\documentstyle[amssymb,a4,12pt,cite]{article}

\newcommand{\rd}{{\rm d}} 
\newcommand{\E}[1]{{\rm e}^{#1}} 

{\catcode `\@=11 \global\let\AddToReset=\@addtoreset}
\AddToReset{equation}{section}

\newtheorem{Theorem} {Theorem} [section]

\newtheorem{Lemma} [Theorem] {Lemma}

\def\scri{\hbox{${\cal J}$\kern -.645em {\raise
      .57ex\hbox{$\scriptscriptstyle (\ $}}}}
\newcommand{\Scri}{\scri} 
\newcommand{\eq}[1]{(\ref{#1})}
\newcommand{\bh}{{\breve h}}
\newcommand{\be}{\begin{equation}}
\newcommand{\ee}{\end{equation}}
\newcommand{\dtwo}{\Delta_2}

\font\SYMA=msam10 

\def\lrcorner{\hbox{\SYMA y}}

\def\Reals{{\Bbb R}}

\def\Naturals{{\Bbb N}}

\def\cosec{\mathop{\rm cosec}\nolimits}
\newcommand{\dx}[1]{{\mbox{\rm d}#1}}
\newcommand{\N}{\Naturals}
\newcommand{\R}{\Reals}

\newfont{\eufm}{eufm10 scaled\magstep1}

\begin{document}

\title{Uniqueness of the Trautman--Bondi mass}

\author{Piotr T.\ Chru\'sciel\thanks{Supported in part by the Polish
    Research Council and by the Austrian Federal Ministry of Science
and Research.  E--mail: Chrusciel@Univ-Tours.Fr} 
\\
D\'epartement de  Mathematiques \\ 
Facult\'e des Sciences et Techniques \\
Universit\'e de Tours \\
Parc de Grandmont, F-37200 Tours, France\\
\\
Jacek Jezierski\thanks{Partially supported by a grant from R\'egion
Centre. E--mail: jjacekj@fuw.edu.pl} 
\\
Department of Mathematical Methods in Physics \\
University of Warsaw \\
ul. Ho\.za 74, 00-682 Warszawa, Poland\\
\\
Malcolm A.H.\ MacCallum\thanks{E--mail:
M.A.H.MacCallum@qmw.ac.uk} 
\\
School of Mathematical Sciences \\
Queen Mary and Westfield College \\
University of London \\
Mile End Road, London E1 4NS, U.K.}

\maketitle

\begin{abstract}
It is shown that the only functionals, within a natural class, which
 are monotonic in time for all solutions of the vacuum Einstein
 equations admitting a smooth ``piece'' of conformal null infinity
 \Scri, are those depending on the metric only through a specific
 combination of the Bondi `mass aspect' and 
other next--to--leading order terms in the metric.
  Under the extra condition of passive BMS invariance, the
 unique such functional (up to a multiplicative factor) is the
 Trautman--Bondi energy. It is also shown that this energy remains
 well-defined for a wide class of `polyhomogeneous' metrics.
\end{abstract}

\section{Introduction}
Consider a Lagrangian theory of fields $\phi^A$ defined on a
 manifold $M$ 
with a Lagrange function density
\be
\label{E.0}
{\cal L} = {\cal L}[\phi^A,\,\partial_\mu \phi^A,\,\ldots,\,
\partial_{\mu_1} \ldots \partial_{\mu_k} \phi^A]\ , 
\ee for some $k
\in \Naturals$, where $\partial_\mu$ denotes partial differentiation
with respect to $x^\mu$. Suppose further that there exists a function
$t$ on $M$ such that $M$ can be decomposed as $\R \times \Sigma$,
where $\Sigma\equiv\{t=0\}$ is a hypersurface in $M$ and the vector
$\partial/\partial t$ is tangent to the $\R$ factor. The proof of the
Noether theorem, as presented {\em
e.g.\/} in \cite[Section 10.1]{Sexl:Urbantke}, shows that the vector
density
\begin{equation}
\label{E.1}
E^\lambda = X^\mu \sum_{\ell=0}^{k-1}  \phi^A{}_{,\alpha_1\ldots\alpha_\ell\mu}
\sum_{j=0}^{k-\ell-1} (-1)^j
\partial_{\gamma_1}\ldots \partial_{\gamma_j}\left(
{{\partial {\cal L}} \over {\partial
 \phi^A{}_{,\lambda\alpha_1\ldots\alpha_{\ell}{\gamma_1}\ldots{\gamma_j}}}} 
\right)  
 - {\cal L} X^\lambda
\end{equation}
has vanishing divergence, ${E^\lambda }_{,\lambda}=0$, when the fields
$\phi_A$ are sufficiently smooth and satisfy the variational equations
associated with a sufficiently smooth ${\cal
L}$ ({\em cf.\/} also \cite{WaldLee}). (This is in any case easily seen by calculating the divergence of
the right--hand--side of eq. \eq{E.1}.)
Here $ \phi^A{}_{,\alpha_1\ldots\alpha_\ell} =
\partial_{\alpha_1}\ldots\partial_{\alpha_\ell} \phi^A$, and
$X^\mu\partial_\mu=\partial_t$. In first order theories, that is
theories in which ${\cal L}$ depends only upon $\phi^A$ and its first
derivatives, it is customary to define
the total energy associated with the hypersurface $\Sigma$ by
the formula
\begin{equation}
\label{E.2}
E(\Sigma) = \int_\Sigma E^\lambda \dx S_\lambda,
\end{equation}
with $\dx S_\lambda = \partial_\lambda \,\lrcorner \,\dx x^0 \wedge
\ldots \dx x^3$, where $\lrcorner$ denotes contraction\footnote{We
  use the conventions that $\partial_0\,\lrcorner \,\dx x^0 \wedge
\ldots \dx x^3 = \dx x^1\wedge\dx x^2 \wedge \dx x^3$,
$\partial_1\,\lrcorner \, \dx x^1\wedge\dx x^2 \wedge \dx x^3 = \dx
x^2 \wedge \dx x^3$, etc.}.
By extrapolation one can also use \eq{E.2} to define an ``energy'' for
higher order theories. Because of its origin, the right-hand-side of
eq.\ \eq{E.2} will be called the {\em Noether energy} of $\Sigma$,
associated with a Lagrange function $\cal L$ and with the vector field $X$.
Now it is well known that the addition to ${\cal L}$ of
a functional of the form
\be
\label{Yf}
\partial_\lambda(Y^\lambda[\phi^A,\,\partial_\alpha \phi^A,\,\ldots,\,
\partial_{\alpha_1} \ldots \partial_{\alpha_{k-1}} \phi^A])\ , \ee
where $k$ is as in \eq{E.0}, does not affect the field
equations\footnote{Here we adopt the standard point of view, that the
field equations are obtained by requiring the action to be stationary
with respect to all compactly supported variations ({\em cf. e.g.\/}
\cite{McCallumTaub} for a discussion of problems that might arise when
this requirement is not enforced).}. We show in Appendix
\ref{transformation} that such a change of the Lagrange function will
change  $E(\Sigma)$ by a boundary integral:
\be
\label{E.4}
E(\Sigma) \ \longrightarrow \ \hat E(\Sigma) = E(\Sigma) +
\int_{\partial\Sigma}
\Delta
E^{\mu\lambda} dS_{\mu\lambda}\ , \ee where $ S_{\alpha\beta} =
\partial_\alpha\,\lrcorner\,\partial_\beta\,\lrcorner\, \dx x^0 \wedge
\ldots \wedge \dx x^3$, with $\Delta E^{\mu\lambda}$ given by eq.\
\eq{VAR.6}.
If $\partial \Sigma$ is a
``sphere at infinity'' the integral over $\partial \Sigma$ has of
course to be understood by a limiting process.  Unless the boundary
conditions at $\partial \Sigma$ force all such boundary integrals to
give a zero contribution, if one wants to define energy using this
framework one has to have a criterion for choosing a ``best''
functional, within the class of all functionals obtainable in this
way. As discussed in more detail in Section \ref{nonunique}, the
vanishing of such boundary integrals will not occur in several cases
of interest.

Now the concept of energy plays a most important role in the context
of fields which are asymptotically flat in light-like directions.  An
appropriate mathematical framework here is that of spacelike
hypersurfaces which intersect the future null infinity $\Scri^+$ in a
compact cross-section $K$. For such field configurations it is 
widely accepted that the ``correct'' definition of energy of a
gravitating system is that given by Freud \cite{F}, Trautman
\cite{T,Tlectures}, Bondi {\em et al.}\ 
\cite{BBM}, and Sachs \cite{Sachs}, which henceforth will be called
the Trautman--Bondi (TB) energy. (Because of the difficulty of
accessing Refs.  \cite{T,Tlectures} we have included an appendix
(Appendix \ref{Trautman}) which describes those results of 
\cite{T,Tlectures} which are related to the problem at hand. This
appendix, together with the date of publication of \cite{T}, should
make it clear why we are convinced that the name of Trautman should be
associated with the notion of mass in the radiating regime in general
relativity.) There have been various attempts to exhibit a privileged
role of that expression as compared with many alternative ones
(\cite{AshS,AshM,AshBR,Hecht:Nester,bicak:energy,%
Penrose:quasi-localmass,Penrose:Rindler:Ch9,%
Sachs:as,YorkBrown,TW,Geroch:WInicour}, to
quote a few), but the papers known to us have failed, for reasons
sometimes closely related to the ones described above, to give a
completely unambiguous prescription about how to define energy at
\Scri. (We make some more comments about that in Section
\ref{nonunique}, {\em cf.\/} also \cite{Walker:Varenna}.) In this
paper we wish to point out that the TB energy is, up to a
multiplicative constant $\alpha \in \Reals$, the {\em only functional
of the gravitational field}, in a certain natural class of
functionals, which is {\em monotonic in
time for all vacuum field configurations} which admit a (piece of) a
smooth null infinity $\Scri^+$.

We shall also consider a second, somewhat larger, class of functionals,
which contains Hamiltonians that arise in an
appropriate symplectic framework. (It will be seen below that the
functionals one obtains from the integrals \eq{E.2} are quadratic
polynomials of the appropriate Bondi functions and their derivatives;
there is no reason for 
the Hamiltonians to satisfy this restriction.)
In that larger class we describe all 
monotone functionals and then among these the further requirement of passive
super--translation invariance also leads to the TB energy as the unique
expression. The symplectic framework which is
appropriate in the context of radiating fields will be described
elsewhere.

It is natural to ask why the Newman-Penrose constants of motion
\cite{NP2}, or
the logarithmic constants of motion of \cite{ChMS},
 do not occur in 
the conclusions of Theorem \ref{T.2}. These quantities are excluded by
the hypothesis that the boundary integrand $H^{\alpha\beta}$ which
appears in the integrals we consider depends on the coordinates only
through the fields. The Newman-Penrose constants could be obtained as
integrals of the form (\ref{E.1}) ({\em cf.\ e.g.\/} \cite{vdB}) if
explicit $r^{2}$ factors were allowed in
$H^{\alpha\beta}$. Similarly logarithmic constants could occur as
integrals of the form (\ref{E.1}) if explicit $1/\ln r$ or $r^{+i}\ln^{-j} r$
factors were allowed there.

This paper is organized as follows: In Section \ref{nonunique} we
review some results about ``energy expressions'' in general
relativity, and comment on non--uniqueness of those. In Section
\ref{monotone} we find all functionals of the fields induced on \Scri\
by the metric which are monotonic in retarded time, in a large class
of natural functionals.  In Section \ref{BMSinvariance} we analyse
those monotonic functionals which are invariant under passive BMS
super--translations, and prove our claim about uniqueness of the
Trautman--Bondi mass. In Section \ref{generalcuts} we give a
super--translation--invariant formula for the Trautman--Bondi
momentum, for general cuts of \Scri.  In Section \ref{polhmg} we
consider the question of convergence of the Freud superpotential to
the Trautman--Bondi mass for space--times with a polyhomogeneous
\Scri.  Remarkably, we find that because of some integral
cancellations the Freud integral always converges to a ``generalized
Trautman--Bondi'' mass, even for metrics which are polyhomogeneous of
order $1$ ({\em cf.\/} Section \ref{polhmg} for
definitions). In
Section \ref{closing} we briefly discuss the potential extensions of
our results to a Hamiltonian setting. An appendix gives a very short
review of Trautman's contribution to the notion of energy for
radiating metrics, while the remaining four appendices contain some
technical results needed in the body of the paper.

\section{Non--uniqueness of the Noether energy for gravitating systems}
\label{nonunique}
As an example of applicability of eq.\ \eq{E.4}, consider a scalar
field $\phi$ in the Minkowski space--time, with $\Sigma=\{t=0\}$.
Assume that $\phi$ satisfies the rather strong fall--off conditions
\be
\label{E.5}
\mbox{for $(t,x)\in \Sigma$ we have}\quad \partial_{\alpha_1}\ldots
\partial_{\alpha_j} \phi=o(r^{-2}),\ 0\le j \le k-1\ , \ee where $k$
is the integer appearing in \eq{E.0}. In this case the boundary
integral in \eq{E.4} will vanish for all smooth $Y^\mu$'s, as
considered in eq.\ \eq{Yf}. This shows that the eq.\ \eq{E.2} leads to
a well--defined notion of energy on this space of fields (whatever the
Lagrange function ${\cal L}$), as long as the volume integral there
converges. (That will be the case if, {\em e.g.}, ${\cal L}$ has no
linear terms in $\phi$ and its derivatives.) 

Consider, next, the same scalar field in Minkowski space--time, with
$\Sigma$ being a hyperboloid, $t=\sqrt{1+x^2+y^2+z^2}$. Suppose further
that ${\cal L}=\nabla^\mu\phi\nabla_\mu\phi$, so that the field equations read
\be\label{E.6}
\Box \phi = 0 \ .
\ee
In that case the imposition of the boundary condition \eq{E.5} does
not seem to be of interest, as such boundary conditions would be
incompatible with the asymptotic behaviour of those solutions of eq.\
\eq{E.6} which are obtained by evolving compactly supported data on
$\{t=0\}$. Thus, even for scalar fields in Minkowski space--time, a
supplementary condition singling out a preferred $E^\lambda$ is needed.

Now for various field theories on the Minkowski background, including
the scalar field, one can impose some further conditions on
$E^\lambda$ which render it unique \cite{Fock,bicak:energy}. The
extension of that analysis to the gravitational field carried on in
\cite{bicak:energy} also leads to a unique $E^\lambda$ (namely the one
obtained from the so--called ``Einstein energy--momentum
pseudo--tensor''), within the class of objects considered. While this is
certainly an interesting observation, the hypotheses made in that last
paper are, however, much more restrictive than is
desirable. It seems therefore that for gravitating systems another
approach is needed. Let us recall how the ``Noether charge'' formalism
described in the Introduction works in that case. There exist various
variational approaches to general relativity, and depending upon the
point of view adopted one finds the following:
\begin{enumerate}
\item 
Let ${\cal L}=\sqrt{|\det g|} R/16\pi$, where the Ricci scalar is considered as
a functional of the metric field $g_{\mu\nu}$, a symmetric connection
$\Gamma^{\alpha}_{\beta\gamma}$, and its first derivatives. In that
case \cite{FerrarisFrancaviglia} one finds
\begin{eqnarray}
E(\Sigma)& = & {1\over 16\pi}\int_\Sigma
\nabla_\mu\nabla^{[\mu}X^{\lambda]}\sqrt{|\det g|}dS_\lambda
\nonumber
\\
& = & {1\over 8\pi}\int_{\partial \Sigma}\nabla^{[\mu}X^{\lambda]}\sqrt{|\det
g|}dS_{\mu\lambda}\ .\label{Komar}
\end{eqnarray}
This integral is known as the Komar energy, except that \eq{Komar} is
actually half of the expression given by Komar \cite{Komar}.
\item
Let ${\cal L}=\sqrt{|\det R_{\mu\nu}|}/\lambda$, where the Ricci tensor
is considered as a functional of a symmetric connection
$\Gamma^{\alpha}_{\beta\gamma}$ and its first derivatives, and
$\lambda $ is a constant. The variational equations for such a theory
are the Einstein equations with a cosmological constant
\cite{Kijowskiold}. The Noether energy gives again \cite{Kijowskiold} the
Komar integral \eq{Komar}.
\item Let ${\cal L}=\sqrt{|\det g|} R/ 16 \pi$, where the Ricci scalar is
considered as a functional of the metric field $g_{\mu\nu}$ and its first
and second derivatives. In that case the value of $E(\Sigma)$ is given
again \cite{Wald:Iyer} by the Komar integral
\eq{Komar} (with a ``wrong'' $1/2$ multiplicative factor).
\item Let ${\cal L}={\cal L}(g_{\mu\nu},g_{\mu\nu,\sigma})$ be the
Einstein Lagrange function \cite{Einstein}, which is obtained by
adding an appropriate divergence to the Hilbert Lagrange function
$\sqrt{|\det g|} R/ 16 \pi$. In that case one obtains \cite{F}
\be
\label{Freud}
E(\Sigma)=\int_{\partial\Sigma} H^{\mu\nu} dS_{\mu\nu}\ ,
\ee
where $H^{\mu\nu}$ is the ``Freud superpotential'' for the ``Einstein
energy--momentum pseudotensor'', {\em cf.\/} eq.\ \eq{Freud2} below.  
\end{enumerate}
Yet another approach, leading to a different
energy expression, can be found in \cite{NovotnyStolin}.

Consider first initial data for, say
vacuum, Einstein equations satisfying the usual fall--off conditions
at spatial infinity;
\be
\label{AF}
g_{\mu\nu}-\eta_{\mu\nu} = O(r^{-1}),\quad \partial_\sigma g_{\mu\nu}=
O(r^{-2})\ .  \ee In that case both the integrals \eq{Komar} and
\eq{Freud} converge.  When the integral over $\partial \Sigma$ in
\eq{Komar} is evaluated on a ``two--sphere at infinity'' in
Schwarzschild space--time one obtains $m/2$. On the other hand, under
the asymptotic conditions \eq{AF} the integral \eq{Freud} coincides with
the standard Arnowitt--Deser--Misner (ADM) expression for energy, and gives
$m$ for that same sphere in Schwarzschild space--time. 

Under the asymptotic conditions \eq{AF}, a way to obtain a unique
expression is given by the symplectic formalism. Namely, one can
require that $E(\Sigma)$ be a Hamiltonian on an appropriately-defined
phase space ({\em cf.\ e.g.\/} \cite{RT,BOM,ChAIHP,KijowskiGRG}). This
requirement, together with the normalization condition that the
Hamiltonian vanishes on Minkowski space--time, uniquely singles out
the Freud--ADM energy as the ``correct'' global energy for general
relativistic initial data sets which satisfy the ``spatial infinity
asymptotic flatness conditions''. Thus the Hamiltonian analysis gives
a rather satisfactory way of singling out an energy expression at
spatial infinity.

Consider, next, hypersurfaces $\Sigma$ which extend to \Scri\ and
intersect \Scri\ transversally.  There have been attempts to use
symplectic methods to define energy in this context
\cite{AshS,AshM,AshBR} (see also
\cite{Helfer,JJAPP98,dkl}). In particular, the
analysis of \cite{AshS,AshM,AshBR} shows that, under appropriate
assumptions, the integral of the time-derivative of the TB energy over
the retarded time gives a Hamiltonian with respect to a proposed
symplectic structure. This does not allow one to extract the integrand
itself from the expression for the Hamiltonian in any unambiguous way,
for reasons somewhat analogous to those described in the Introduction.
Moreover in those papers one has to assume various decay properties of
the fields on $\Scri$ for large absolute values of the retarded time,
which have not been established so far. Finally, as the symplectic
structure considered in \cite{AshS,AshM,AshBR} has a perhaps less
universally accepted status than the one considered on standard
asymptotically flat hypersurfaces, one should perhaps also face the
question of uniqueness of the symplectic structures involved. For all
those reasons we conclude that the framework of \cite{AshS,AshM,AshBR}
fails to demonstrate uniqueness of the TB mass.

\section{Monotonic functionals}\label{monotone}
From now on, we shall consider metrics $g_{\alpha\beta}$ defined on
appropriately large subsets of $\Reals^4$, but not necessarily
globally defined on
$\Reals^4$, and satisfying Einstein's equations near \Scri. We
shall examine
a class of functionals which includes all the cases discussed in section
\ref{nonunique}, and in particular all functionals differing from the
Hilbert Lagrangian by a divergence. These functionals have the form
\begin{eqnarray}
\label{G.1}
H[u_0,\,g] &=&
\lim_{\rho\to\infty}\int_{S(t=u_0+\rho,\rho)} H^{\alpha\beta}[g] \, \dx
S_{\alpha\beta}, \\ 
\dx S_{\alpha\beta} &=&
\partial_\alpha\,\lrcorner\,\partial_\beta\,\lrcorner\, \dx x^0 \wedge
\ldots \wedge \dx x^3, \nonumber
\end{eqnarray}
where
\begin{equation}
\label{G.1.1}
H^{\alpha\beta}[g]{(x)} \equiv
H^{\alpha\beta}(g_{\mu\nu}{(x)},\,
\partial_\alpha g_{\mu\nu}{(x)},\,\ldots,\,\partial_{\alpha_1}\ldots
\partial_{\alpha_k} g_{\mu\nu}{(x)})\ .
\end{equation}
for some $k \in \Naturals$,
and $H^{\alpha\beta}$ is a twice continuously differentiable function
of its arguments. Here $S(\tau,\,\rho)$ denotes a sphere
$r\equiv\sqrt{(x^1)^2+(x^2)^2+(x^3)^2} = \rho$, $t\equiv x^0=\tau$. 
The metrics
$g_{\alpha\beta}$ will be assumed to satisfy the standard fall-off
conditions corresponding to asymptotic flatness at null
infinity. More precisely, consider a space-time $(M,\,g)$ which
admits a conformal completion (which in this section we consider to be
smooth) in the following sense: there exists a manifold with boundary
$(\bar{M},\,\bar{g})$, a diffeomorphic embedding $\Phi: M \rightarrow
\bar{M}\setminus \partial \bar{M}$, and a smooth function $\Omega$ on
$\bar{M}$ such that $\Phi^{\ast}(\Omega^{-2}\bar{g}) = g$. We shall
also assume that $\Omega_{|\partial \bar{M}} = 0$, that $\dx \Omega$
is nowhere vanishing on $\partial \bar{M}$, and that $\Scri \equiv
\partial \bar{M}$ is diffeomorphic to $I \times S^2$ where $I$ is an
interval (possibly but not necessarily equal to $\Reals$). By a
standard construction we can introduce Bondi coordinates near $\Scri$
({\em cf.\ e.g.\/} \cite{TW} or \cite{ChMS}) so
that we have
\begin{equation}
\label{G.2}
\dx s^2 = -\frac{V e^{2\beta}}{r}\, \dx u^2 - 2e^{2\beta}\, \dx u\,
\dx r
+ r^2\, h_{ab} (\dx x^a - U^a \dx u) (\dx x^b - U^b \dx u)\,,
\end{equation}
with $x^a=(\theta,\,\phi)$. We can introduce quasi-Minkowskian
coordinates by setting
\begin{equation}
\label{G.3}
u=t-r,\quad x=r \sin \theta \cos \phi, \quad  y=r \sin \theta \sin
\phi, \quad  z=r \cos \theta.
\end{equation}

We shall consider only vacuum metrics; recall that this implies the
following behavior of $h_{ab}$, $\beta$, $U^a$ and $V$
(\cite{vdB,BBM,Sachs})
\begin{eqnarray}
h_{ab} &=& \bh_{ab}\left(1+\frac 1{4r^2}\chi^{cd}\chi_{cd}\right) +
\frac{\chi_{ab}(v)}{r} + O(r^{-3})\ , 
\nonumber \\
\beta &=& -\frac{1}{32} \frac{\bh^{ab}\bh^{cd}\chi_{ac}\chi_{bd}}{r^2}
+O(r^{-3})\ , \nonumber\\
\label{G.4}
U^a &=& -\frac{1}{2} \frac{{\cal D}_b \chi^{ab}}{r^2} +
\frac{32N^a(v)+{\cal D}^a\left(\chi^{cd}\chi_{cd} \right) +8
\chi^a{_b}{\cal D}_c \chi^{bc}}{16r^3}
+O(r^{-4}),\\ 
V &=& r-2M(v) +\frac{\chi^{cd}\chi_{cd}+4{\cal D}_b\chi^{ab}
{\cal D}^c\chi_{ac}-16{\cal D}_a N^a }{16r} + O(r^{-2})\ ,\nonumber
\end{eqnarray}
Here $(v) \equiv (u,\, x^a)$ and $\bh_{ab} \dx x^a \dx x^b = \dx
\theta^2 +\sin^2 \theta\,\dx \phi^2$; ${\cal D}_a$ is the covariant
derivative operator defined by $\bh_{ab}$. Indices $a$, $b$, {\em
etc.}, take values
 2 and 3, and 
 are raised and lowered with $\bh^{ab}$. The tensor field $\chi_{ab}$
 satisfies the condition
\begin{equation}
\label{G.5}
\bh^{ab}\chi_{ab} = 0,
\end{equation}
and no other conditions are imposed\footnote{\label{f1}Note,
  however, that there may be some restrictions arising from some
  further global hypotheses if those are made. We emphasize that we do
  not impose any such global hypotheses here.} on $\chi_{ab}(v)$ by the vacuum
Einstein equations.  The functions $M$ and $N^a$
satisfy the following equations
\begin{eqnarray}
\frac{\partial M}{\partial u} &=& -\frac18
\bh^{ac}\bh^{bd}\dot\chi_{ab}\dot\chi_{cd} + \frac14
{\cal D}_a {\cal D}_b \dot\chi^{ab} \ ,\nonumber \\
\label{G.6}
3\frac{\partial N^a}{\partial u} &=& -{\cal D}^a M +\frac14 \epsilon^{ab} {\cal
D}_b \lambda - K^a\ , \\
K^a &\equiv & \frac34\chi^a{_b}{\cal D}_c\dot\chi^{bc} +
\frac14\dot\chi^{cd} {\cal D}_d{\chi^{a}}_{c}\ ,
\nonumber \\ 
\lambda &\equiv & \bh^{bd} \epsilon^{ac}{\cal D}_c {\cal D}_b \chi_{da}
 \ ,\nonumber
\end{eqnarray}
{where $\dot\chi:=\partial_u\chi$}. 
Here $\epsilon_{ab}=\partial_a \lrcorner \partial_b \lrcorner
\dx{}^2\mu$ where
$\dx{}^2\mu =\sin \theta\,\dx \theta \wedge \dx\phi = \frac{1}{2}
\epsilon_{ab}dx^a\wedge dx^b$ is the standard volume form on
$S^2$.
If we fix some $u_0 \in I$, then
the Einstein equations do not impose$^{\mbox{\scriptsize\ref{f1}}}$ any
restrictions on the function  
$M(u_0,\,\theta,\,\phi)$ and the vector field
$N^a(u_0,\,\theta,\,\phi)$ on $S^2$.

Equation (\ref{G.4}) shows that in the coordinate system (\ref{G.3})
the metric (\ref{G.2}) is of the form
\begin{equation}
\label{G.6.1}
g_{\mu\nu} = g^0_{\mu\nu}+
\frac{g^1_{\mu\nu}(v)}{r}+\frac{g^2_{\mu\nu}(v)}{r^2} +O(r^{-3}),
\end{equation}
with obvious analogous expansions holding for the various
derivatives of $g_{\mu\nu}$ when an appropriate expansion for the
derivatives of $h_{ab}$ is assumed. Here $g^0_{\mu\nu} =
\mbox{\rm diag}(-1,\,1,\,1,\,1)$. We can now insert a metric of the form
(\ref{G.6.1}) into a functional of the form (\ref{G.1}), and as a
further restriction we shall require that $H$ has a finite numerical
value for all fields $g_{\mu\nu}$ of the type described above. Our
hypothesis of differentiability of $H^{\alpha\beta}$ allows us to
Taylor expand  $H^{\alpha\beta}$ to order $2$ in terms of powers of
$g_{\mu\nu}-g^0_{\mu\nu}$,
$\partial_\sigma(g_{\mu\nu}-g^0_{\mu\nu})=\partial_\sigma g_{\mu\nu}$,
{\em etc.},
about $g_{\mu\nu}=g^0_{\mu\nu}$. Note that by (\ref{G.1.1}) the
$H^{\alpha\beta}[g^0_{\mu\nu},\,0,\,\ldots,\,0]$ are constants which
are either zero or integrate to zero in (\ref{G.1}) (otherwise the
limit in (\ref{G.1}) would be infinite), so that
$$ H[u,\,g^0_{\mu\nu}] = 0 \qquad \forall u \in I .$$ The $1/r$ terms
in $g_{\mu\nu}$ and its $v$ derivatives will give at most a quadratic
contribution to $H$, and the $1/r^2$ terms at most a linear one, while
the remainder terms in the Taylor expansion of $H^{\alpha\beta}$ will
contribute nothing in
the limit $r \rightarrow \infty$. It follows that $H$ can be written
in the form
\begin{equation}
\label{G.7}
H = \int_{S^2}
h[M,\,M^{(1)},\,\ldots,\,M^{(k)},\,N^a,\,N^{a(1)},\,\ldots,\,N^{a(k)},\,
\chi_{ab},\,\chi_{ab}^{(1)},\,\ldots\,\chi_{ab}^{(k)},\theta,\phi]\dx{}^2\mu\ .
\end{equation}
 Here the
addition of a superscript $(\ell)$ to a quantity denotes the $\ell$-th
$u$-derivative of that quantity. The square brackets around the
arguments of $h$ are meant to emphasize the fact that $h$ is not a
function but a local functional of the fields which is a
differentiable function of $M$, $\partial_\alpha M$, $\ldots$,
$\partial_{\alpha_1} \ldots \partial_{\alpha_k} M$, $M^{(1)}$,
$\partial_\alpha M^{(1)}$, $\ldots$, $\partial_{\alpha_1} \ldots
\partial_{\alpha_k} M^{(1)}$, $\ldots$, etc., for some finite number
of derivatives in directions tangent to $S^2$. Note that for
functionals (\ref{G.1}) the dependence of $H$ on $N^a$, $N^{a(1)}$,
$\ldots$, {\em etc.}, as well as on derivatives of $N^a$, $N^{a(1)}$,
$\ldots$, {\em etc.}, in angular directions, will be linear because
$N^a$ comes with a factor $r^{-2}$, and we shall henceforth only
consider such functionals. {}From a symplectic point of view it turns
out to be natural not to make the hypothesis that $h$ is a quadratic
polynomial of the fields and their derivatives, as would be the case
for functionals (\ref{G.1}), and for this reason we shall allow
arbitrary differentiable functions $h$ in
(\ref{G.7}), 
 except for the hypothesis of linearity in $N^a$ with
coefficients of linearity independent of the remaining fields.

Assuming that the metric $g_{\mu\nu}$ is vacuum (at least in a
neighborhood of \Scri) we can eliminate the $u$-derivatives of $M$
and $N^a$ in favour of $M$ and $u$-derivatives of $\chi_{ab}$, using
equations (\ref{G.6}), so that (\ref{G.7}) can be rewritten as
\begin{equation}
\label{G.8}
H = \int_{S^2}
h[M,\,N^a,\,\chi_{ab},\,\chi_{ab}^{(1)},\,\ldots\,\chi_{ab}^{(k)}]
\dx{}^2\mu\ ,
\end{equation}
where $h$ is still linear in $N^a$ and its derivatives in directions
tangent to $S^2$. By an abuse of notation here, we still denote the
integrand of $H$ by $h$, although it will in general not coincide with
the original $h$ of (\ref{G.7}). Before we proceed further we need the
following result based on the work of Friedrich \cite{F1,F2} as
extended by Kannar \cite{K}:
\begin{Lemma}
\label{L.1}
Let $M_0(\theta,\,\phi)$ be a smooth function on $S^2$,
$N_0^a(\theta,\,\phi)$ be a smooth vector field on $S^2$, and
$\chi^0_{ab}(\theta,\,\phi)$, $\chi^1_{ab}(\theta,\,\phi)$, \ldots,
$\chi^k_{ab}(\theta,\,\phi)$, \ldots be any sequence of smooth
symmetric traceless tensors on $S^2$. Then there exists $\epsilon >0 $
and a vacuum space-time $(M,\,g)$ with a smooth conformal completion
(in the sense described above) $(\bar{M},\,\bar{g})$ which has a
spherical cut $u=u_0$ of $\Scri^+\supset
(u_0-\epsilon,u_0)\times S^2$ such
that the Bondi functions $M(u,\,\theta,\,\phi)$,
$N^a(u,\,\theta,\,\phi)$ and $\chi_{ab}(u,\,\theta,\,\phi)$ satisfy
\begin{eqnarray}
\lim_{u\to u_0}
M(u,\,\theta,\,\phi) &=& M_0(\theta,\,\phi)\ , \label{PE.-1} \\
\label{PE.0}
\lim_{u\to u_0}
N^a(u,\,\theta,\,\phi) &=& N^a_0(\theta,\,\phi)\ , \\
\label{PE.1}
\quad \forall i \in \Naturals \qquad\lim_{u\to u_0}
\chi^{(i)}_{ab}(u,\,\theta,\,\phi) &=&
\chi^i_{ab}(\theta,\,\phi)\  .
\end{eqnarray}
\end{Lemma}

\noindent{\bf Remarks:}
1. Actually the functions $\chi_{ab}(u,\,\theta,\,\phi)$ can be
arbitrarily  prescribed as functions of $(u,\,\theta,\,\phi)$ on an
interval $(u_0-\epsilon,\,u_0]$ for some appropriate $\epsilon$. The
above weaker claim is, however, sufficient for our purposes.

2. The limits $\lim_{u\to u_0}$ in the equations above have been
introduced to avoid talking about space--times with
boundary.

\vspace{\baselineskip}

\noindent{\bf Proof:}
For 
$x\in [0,1]$
and $u_0-1 \leq u \leq u_0$ consider metrics of the form
\begin{equation}
\label{PE.3}
\bar{g}_{\mu\nu}\dx x^\mu \dx x^\nu = -Vx^3e^{2\beta}\, \dx u^2 +
2e^{2\beta}\, \dx u\,\dx x
+h_{ab} (\dx x^a - U^a \dx u) (\dx x^b - U^b \dx u)\,,
\end{equation}
We wish to show that we can find $\epsilon > 0$ and a metric
$\bar{g}_{\mu\nu}$ of the form (\ref{PE.3}) defined for $x \in
[0,\,\epsilon]$, $x^a \in S^2$, $u \in (u_0-\epsilon,\,u_0]$ such that
$x^{-2} \bar{g}_{\mu\nu}$ satisfies the Einstein vacuum equations, for
which \eq{PE.-1}--\eq{PE.1} hold.  We
shall construct the appropriate solution backwards in $u$ on
$(u_0-\epsilon,\,u_0] \times [0,\,\epsilon] \times S^2$ by solving an
asymptotic characteristic initial value problem with data given on the
null hypersurface ${\cal N} = \{u=u_0\}$ and on a piece of \Scri$=\{x=0, u
\in (u_0-1,\,u_0]\}\times S^2$. This proceeds as follows:
For $0 \leq x \leq 1$ let $h^0_{ab}(x,\,\theta,\,\phi)$ be any
$x$-dependent family of symmetric non-degenerate tensors on $S^2$ with
$h^0_{ab}(0,\,\theta,\,\phi)=\bh_{ab}(\theta,\,\phi)$ (the standard
metric on $S^2$), with $\det h^0_{ab}(x,\,\theta,\,\phi)=\det
\bh_{ab}(\theta,\,\phi)$,  and with 
\begin{equation}
\label{PE.2}
\left. \frac{\partial h^0_{ab}}{\partial x} \right|_{x=0} =
\chi^0_{ab} 
\ .
\end{equation}
set $h_{ab}(u=u_0,x,\theta,\phi) = h_{ab}^0(x,\theta,\phi)$.
Using the Bondi--van der Burg--Sachs prescription in the coordinate
system $(r=1/x,\,x^a)$ \cite{BBM,Sachs} we can find unique smooth
functions $\partial^i_u \beta(u_0,\,x,\,\theta,\,\phi)$, $\partial^i_u
U^a(u_0,\,x,\,\theta,\,\phi)$, $\partial^i_u
V(u_0,\,x,\,\theta,\,\phi)$, $\partial^i_u
h_{ab}(u_0,\,x,\,\theta,\,\phi)$, such that eqs.\
\eq{PE.-1}--\eq{PE.1} hold and such that for all $N\in \N$ the metric
$g_{\mu\nu}=x^{-2}\bar{g}_{\mu\nu}$ satisfies $R_{\mu\nu} =
O((u-u_0)^N)$, whatever the fields $\beta,V,U,h_{ab}$ as long as those
fields and their derivatives assume the boundary values obtained
above. Indeed, the fields appearing at
the right hand side of eqs. \eq{PE.-1}--\eq{PE.1} provide precisely
the data needed for the construction of a solution of the hierarchy of
equations obtained by $u$-differentiating the Bondi--van der
Burg--Sachs equations.  It follows that any geometric quantities,
built out of the metric together with an arbitrary finite number of
its derivatives, calculated at $u=u_0$ for any two such metrics will
coincide at $u=u_0$.

Because $\chi_{ab}$ is symmetric and traceless, $h_{ab}$ can be parametrized
as
\begin{equation}
\label{PE.4}
\left( \begin{array}{cc}
e^{2\gamma} \cosh (2\delta) & \sinh (2\delta) \sin \theta \\
\sinh (2\delta) \sin \theta & e^{-2\gamma} \cosh (2\delta) \sin^2 \theta
\end{array} \right)\ ,
\end{equation}
where we write $\gamma = c(v)/r +O(r^{-3})$, $\delta=d(v)/r
+O(r^{-3})$.
Let $e_{AB'}$ be the following tetrad field
\begin{eqnarray}
e_0 &=& e_{00'} = -\partial_x\ , \nonumber \\
\label{PE.3.1}
e_1 &=& e_{11'} = e^{-2\beta}(\partial_u + \frac{1}{2}Vx^3\partial_x
 +U^\theta\partial_\theta+U^\phi \partial_\phi) \ ,  \\
e_2 &=& e_{01'} = \frac{1}{\sqrt{2}}(e^{-\gamma}
                    (-\cosh \delta +i \sinh \delta) \partial_\theta
 + e^\gamma (\sinh \delta -i \cosh \delta) \cosec \theta \partial_\phi)
 \ ,\nonumber\\
e_3 &=& e_{10'} = (e_2)^{\ast}\ ,\nonumber
\end{eqnarray}
where $(e_2)^{\ast} $ denotes the vector whose coordinate components
are complex conjugates of those of $e_2$. From
(\ref{PE.4})--(\ref{PE.3.1}) one can calculate the Newman-Penrose
quantity $\sigma$ ($=\Gamma_{01'00}$ in the notation of \cite{K}, and
$=\Gamma_{0001'}$ in the notation of \cite{Newman:Penrose}) to obtain
\begin{equation}
\label{PE.5}
\sigma |_{x=0} = c +id\ .
\end{equation}
In \cite{K} the time sense of $e_0$ and $e_1$ is unspecified and $e_2$
is only specified up to rotations in the  $e_2$--$e_3$ plane at points
in the intersection of N and \Scri\ (in Kannar's notation). Since they
are then  parallelly propagated in Kannar's treatment, these
rotations are $u$ and $r$ independent. Note that up to
these ambiguities the tetrad (\ref{PE.3.1}) coincides with that used
in \cite{K}, time-reversed, at $x=0$, but will in general differ from
it at other points. This is irrelevant as far as the value of $\sigma
|_{x=0}$ is concerned because $\sigma$ at $x=0$ is calculated using
only derivatives of the tetrad field tangent to the spheres $x=0$, $u
=$ constant, so that $\sigma |_{x=0}$ calculated for the tetrad
(\ref{PE.3.1}) will coincide with that calculated in the tetrad used
in \cite{K} (up to a constant factor of modulus 1). The essential
point is that $c$ and $d$ give the requisite data. 

Now let $\chi_{ab}(u,\,\theta,\,\phi)$ be an arbitrary one-parameter
family of symmetric tensor fields on $S^2$, with $u \in (u_0-1,\,u_0]$
such that $\partial^i_u \chi_{ab}(u_0,\,\theta,\,\phi)=
\chi^i_{ab}(\theta,\,\phi)$; from $\chi_{ab}(u,\,\theta,\,\phi)$ we
can calculate $\sigma |_{x=0}$. 
{}From $\partial^i_u\bar{g}_{\mu\nu}|_{u=u_0}$, $i=0,1,2$
(which we have
already calculated previously)
we can determine at $u=u_0$ the remaining initial
data needed for the Friedrich--Kannar asymptotic initial value problem.
The existence of an $\epsilon > 0$  (depending on the initial
data) and a solution of the vacuum Einstein equations
${g}_{\mu\nu}$ defined for $(u,x,\theta,\phi)\in(u_0-\epsilon,\,u_0]
\times [0,\epsilon] \times S^2$ 
assuming those initial data now follows from the main theorem of
\cite{K}. The property that the Bondi functions 
${M}$,
${N}^a$, ${h}_{ab}$ and $\partial_x{h}_{ab}$ parametrizing
the metric 
$\bar{g}_{\mu\nu}$ assume the desired values 
on \Scri$^+$ follows from the uniqueness theorems of \cite{F1}.
\hfill $\Box$

We can now pass to the proof of our main result.
\begin{Theorem}
\label{T.1}
Consider any functional of the form
\begin{eqnarray}
H[g] &=& \int_{S^2} [h(M,\,{\cal D}_a M,\,\ldots,\,{\cal
 D}_{a_1} \ldots {\cal D}_{a_k} M,  
\nonumber \\
&& \phantom{\int_{S^2} [h(}
 \chi_{ab},\,{\cal D}_c \chi_{ab},\,\ldots,\,{\cal
 D}_{c_1} \ldots {\cal D}_{c_k} \chi_{ab}, \nonumber \\
&& \phantom{\int_{S^2} [h(}
 \chi^{(1)}_{ab},\,{\cal D}_c \chi^{(1)}_{ab},\,\ldots,\,{\cal
 D}_{c_1} \ldots {\cal D}_{c_k} \chi^{(1)}_{ab}, \nonumber \\
&& \phantom{\int_{S^2} [h(}
 ,\,\ldots,\,\nonumber \\
&& \phantom{\int_{S^2} [h(}
 \chi^{(k)}_{ab},\,{\cal D}_c \chi^{(k)}_{ab},\,\ldots,\,{\cal
 D}_{c_1} \ldots {\cal D}_{c_k} \chi^{(k)}_{ab},\,x^a) \nonumber
 \\
\label{E2.1}
&& \phantom{\int_{S^2} [}
+ \alpha_a N^a +\alpha_{ab}{\cal D}^a N^b + \ldots +\alpha_{a_1\ldots
a_k b}{\cal D}^{a_1} \ldots {\cal D}^{a_k} N^b] \dx{}^2\mu,
\end{eqnarray}
where $h$ is a twice continuously differentiable function of all its
arguments, with some, say smooth, tensor fields $\alpha_{a_1\ldots a_k
  b}$ on $S^2$.
If $H$ is monotone non-increasing in $u$ for all metrics $g$ which
satisfy the vacuum Einstein field equations (with $M$, $\chi_{ab} $
and $N$  interpreted as Bondi functions appearing in $g$), then $H$
can be rewritten as
$$H=\int_{S^2} \Psi(M-\frac{1}{4}\bh^{ac}\bh^{bd}{\cal D}_a{\cal D}_b
\chi_{cd},\, x^a)
\dx{}^2\mu,$$ 
with a differentiable local functional\footnote{That is,  
$\Psi(f)$ is a differentiable function of $x^a$, $f$ and a finite number of
  its derivatives in directions tangent to $S^2$.}
$\Psi(f)$ whose
variational derivative $\delta \Psi/\delta f$ is  non--negative.
\end{Theorem}

\noindent{\bf Proof:}
Note first that the tensor fields $\alpha_{ab}$, \ldots,
$\alpha_{a_1\ldots a_k b}$, can be set to zero by integration by parts
and a redefinition of $\alpha_{b}$,  $\alpha_{b} \rightarrow
\hat{\alpha}_{b}$ with an appropriate  $\hat{\alpha}_{b}$. Calculating
the $u$-derivative of (\ref{E2.1}) we obtain
\begin{equation}
\label{E2.2}
\frac{\dx H}{\dx u} = \int_{S^2} (
\frac{\delta h}{\delta 
M}\dot{M} + \hat{\alpha}_a\dot{N}^a \nonumber \\
+  \frac{\delta h}{\delta \chi_{ab}}\dot{\chi}_{ab} + \ldots +
  \frac{\delta h}{\delta \chi^{(k)}_{ab}}\chi^{(k+1)}_{ab}
  )\dx{}^2\mu\ .
\end{equation}
Now $\delta h / \delta \chi^{(k)}_{ab}$ and all the terms in
(\ref{E2.2}) except for the last one are independent of $\chi^{(k+1)}(u_0)$.
If  $\delta h / \delta
\chi^{(k)}_{ab}$ were non-zero for some $k \geq 1$ we could, by Lemma
\ref{L.1}, find a solution of the vacuum Einstein equations with
$\chi^{(k+1)}(u_0)$ so chosen that $\dx H/\dx u > 0$, which shows that
 $\delta h / \delta \chi^{(k)}_{ab}=0$ for all $k \geq 1$. Setting
\begin{eqnarray*}
 \hat{h}[M,\,\chi_{ab},\,x^a] &=&
 h(M,\,{\cal D}_a M,\,\ldots,\,{\cal D}_{a_1} \ldots {\cal D}_{a_k} M, \\
&& \phantom{h(}
 \chi_{ab},\,{\cal D}_c \chi_{ab},\,\ldots,\,{\cal D}_{c_1} \ldots
 {\cal D}_{c_k} \chi_{ab}, \\
&& \phantom{h(}
 \chi^{(1)}_{ab}=0,\,{\cal D}_c \chi^{(1)}_{ab}=0,\,\ldots,\,{\cal
 D}_{c_1} \ldots {\cal D}_{c_k} \chi^{(1)}_{ab}=0, \nonumber \\
&& \phantom{h(}
 ,\,\ldots,\,
 \chi^{(k)}_{ab}=0,\,{\cal D}_c \chi^{(k)}_{ab}=0,\,\ldots,\,{\cal
 D}_{c_1} \ldots {\cal D}_{c_k} \chi^{(k)}_{ab}=0,\, x^a)\ ,
\end{eqnarray*}
we obtain from  $\delta h / \delta \chi^{(k)}_{ab}=0$, $k \geq 1$,
\begin{eqnarray}
H(g,\,u) &=& \int_{S^2} (\hat{h}[M,\,\chi_{ab}] +
\hat{\alpha}_aN^a)\dx{}^2\mu, \nonumber \\
\label{E2.3}
\frac{\dx H}{\dx u} &=& \int_{S^2} (
\frac{\delta \hat{h}}{\delta
M}\dot{M} + \hat{\alpha}_a\dot{N}^a + \frac{\delta \hat{h}}{\delta
\chi_{ab}}\dot{\chi}_{ab})\dx{}^2\mu. 
\end{eqnarray}
Consider, first, equation (\ref{E2.3}) for solutions of the vacuum
Einstein equations with $\dot{\chi}_{ab}|_{u=u_0}=0$. Equations
(\ref{E2.3}) and (\ref{G.6}) then yield
\begin{eqnarray}
\frac{\dx H}{\dx u} &=& \int_{S^2} (
- \frac13\hat{\alpha}_a({\cal
D}^aM -\frac14\epsilon^{ab}{\cal D}_b\lambda))\dx{}^2\mu \nonumber \\
\label{E2.4}
&=& \int_{S^2} (
\frac13 M{\cal
D}^a \hat{\alpha}_a - \frac1{12}\lambda\epsilon^{ab}{\cal D}_b
\hat{\alpha}_a )\dx{}^2\mu\ .
\end{eqnarray}
To proceed further we need to know a little more about $\lambda$ as
defined by (\ref{G.6}).  In Appendix \ref{AppA} we show that the
image of the operator $\chi_{ab} \rightarrow \epsilon_{ac}{\cal D}^c
{\cal D}_b \chi^{ab}$ defined on traceless symmetric tensors consists
precisely of functions of the form $P\psi$, where $\psi$ is an
arbitrary appropriately differentiable function on $S^2$ and $P$ is
the projection operator defined as
\begin{equation}
\label{M.1}
P\psi = \psi - \sum_{i=0}^3 \Phi_i \int_{S^2} \psi \Phi_i \dx{}^2 \mu\ 
,
\end{equation}
where the $\Phi_i$ form an orthonormal basis of the space of spherical
harmonics with $\ell = 0$ ($\Phi_0$) and $\ell = 1$ ($\Phi_i$,
$i=1,\,2,\,3$). Consequently $\lambda$ runs over all smooth functions
with no zero or first spherical harmonics as $\chi_{ab}$ runs over
all smooth symmetric traceless tensors. This, together with Lemma
\ref{L.1} (note that $M$ in \eq{E2.4} is arbitrary) shows that $\dx
H/\dx u$ in (\ref{E2.4}) will have an arbitrary sign unless \be
\label{M.1.1}{\cal D}_a \hat{\alpha}^a =0, \qquad \epsilon^{ab} {\cal D}_a
\hat{\alpha}_b = \sum_{i=1}^3 \alpha_i \Phi_i\ ,
\ee 
for some constants $\alpha_i$.  It follows  that 
\begin{equation}
\label{E2.5}
\hat{\alpha}^a = \frac12{\cal D}_b(\epsilon^{ab}\sum_{i=1}^3
\alpha_i\Phi^i)
\end{equation}
(the fact that the above vector field satisfies \eq{M.1.1} can be checked
by a direct calculation; the fact that there is only one such vector
field is shown in Appendix \ref{AppA}).
Returning to equation (\ref{E2.3}), we obtain from (\ref{E2.5}) and
(\ref{G.6})
\begin{eqnarray}
  \frac{\dx H}{\dx u} &=& \int_{S^2} \left[\frac{\delta \hat{h}}{\delta
    M}\left(-\frac18\dot{\chi}_{ab}\dot{\chi}^{ab}+\frac14{\cal D}_a{\cal
    D}_b\dot{\chi}^{ab}\right) -\frac{1}{3}\hat{\alpha}_a K^a + \frac{\delta
    \hat{h}}{\delta \chi_{ab}} \dot{\chi}_{ab}\right]\dx{}^2\mu \nonumber\\ 
\label{E2.6}
&=& \int_{S^2} \left[
\left(\frac14{\cal D}^a{\cal D}^b \frac{\delta \hat{h}}{\delta M}
+\frac{1}{6}\hat{\alpha}_c{\cal D}^b\chi^{ca} +\frac14\chi^{ca}{\cal
D}^b\hat{\alpha}_c +
 \frac{\delta\hat{h}}{\delta \chi_{ab}}\right)\dot{\chi}_{ab}
\right. 
\nonumber \\
&&\phantom{\int_{S^2}\frac{\delta \hat{h}}{\delta M}} \left. -\frac18
\frac{\delta \hat{h}}{\delta M}
\dot{\chi}_{ab}\dot{\chi}^{ab}\right]\dx{}^2\mu \ .
\end{eqnarray}
Define a new functional $\hat{\Psi}$ by
$$\hat{\Psi}[f,\,\chi_{ab},\,x^a] := \hat{h}[M=f+\frac14{\cal D}_a {\cal D}_b
\chi^{ab},\,\chi_{ab},\,x^a].$$
(\ref{E2.6}) can be rewritten  as
\begin{equation}
\frac{\dx H}{\dx u} =
\label{E2.7}
-\frac18 \int_{S^2} \frac{\delta \hat{\Psi}}{\delta f}
\dot{\chi}_{ab}\dot{\chi}^{ab} \dx{}^2\mu 
 +\int_{S^2} \left[
\left(\frac{1}{6}\hat{\alpha}_c{\cal D}^b\chi^{ca} +\frac14\chi^{ca}{\cal
D}^b\hat{\alpha}_c + \frac{\delta\hat{\Psi}}{\delta
\chi_{ab}}\right)\dot{\chi}_{ab}\right]\dx{}^2\mu  \ .
\end{equation}
$\dx H/\dx u$ will be non-positive for all $\dot{\chi}_{ab}$ if and
only if $\delta \hat{\Psi}/\delta f$ is non-negative, and the last
integral vanishes, which yields
\be
\label{addedeq}
\frac{\delta\hat{\Psi}}{\delta
\chi_{ab}}= -  
\left(\frac{1}{6}\hat{\alpha}_c{\cal D}^b\chi^{ca} +\frac14\chi^{ca}{\cal
D}^b\hat{\alpha}_c \right)\ .
\ee
We wish to show that $\hat\alpha^a$ has to be zero. To do this, fix a
smooth $f$ and consider $G_f[\chi_{ab},\,x^a] = \int \hat{\Psi}[f, 
\,\chi_{ab},\,x^a] \dx{}^2 \mu$
as a functional of $\chi_{ab}$. Note that
if we endow the space of the $\chi_{ab}$'s with a Sobolev space
topology $W_{k,2}(S^2)$ with some $k$ large enough, then $G_f$ will be a
twice differentiable function on that space, and by \eq{addedeq} we have
\begin{equation}
\label{E2.8.1}
G_f'[\nu]:= -\frac{1}{12}\int_{S^2}\left(
 2\hat{\alpha}_c{\cal D}^{b}\chi^{ca} +3\chi^{ac}{\cal
 D}^{b}\hat{\alpha}_c \right)\nu_{ab}\dx{}^2\mu\ ,
\end{equation}
where $G_f'[\nu]$ denotes the derivative of $G_f$ acting on the symmetric
traceless tensor $\nu$. It follows from Schwarz's Lemma that the
second derivative $G_f''$ of $G_f$ satisfies $G_f''[\tau,\nu]=
G_f''[\nu,\tau]$, for all smooth symmetric traceless tensor fields
$\tau_{ab}$ and $\nu_{ab}$.  {}From \eq{E2.8.1} we have
\begin{equation}
\label{E2.8}
G_f''[\tau,\nu]:= -\frac{1}{12}\int_{S^2}\left(
 2\hat{\alpha}_c{\cal D}^{b}\tau^{ca} +3\tau^{ac}{\cal
D}^{b}\hat{\alpha}_c \right)\nu_{ab}\dx{}^2\mu.
\end{equation}
Letting $F:= \frac12 \sum_{i=1}^3 \alpha_i\Phi^i$ with some constants
$\alpha_i$, we have 
$\hat\alpha^a=\epsilon^{ab}{\cal D}_b F$ ({\em cf.\/} \eq{E2.5}).
We also have ${\cal D}^a{\cal D}_b F= -\delta^a_b F$
({\em cf.\ e.g.\/} \cite[Lemma 5]{beig:sh}),
so one gets
${\cal D}^a\hat\alpha^b= 
\epsilon^{ab} F$. Using those identities,
by integration by parts one obtains
\[G_f''[\tau,\nu]- G_f''[\nu,\tau]=
-\frac{1}{6}\int_{S^2}\left(
 \hat{\alpha}_c{\cal D}^{b}\tau^{ca} +\hat{\alpha}^b{\cal D}_{c}\tau^{ca} +
 2 F \tau^{ac}\epsilon^{b}{_c} \right)\nu_{ab}\,\dx{}^2\mu\ .
\]
Since $\nu$ is arbitrary (traceless, symmetric) we obtain
\be
\label{muid}
 TS\left[ 
 \hat{\alpha}_c{\cal D}^{b}\tau^{ca} +\hat{\alpha}^b{\cal
 D}_{c}\tau^{ca} +  2 F \tau^{ac}\epsilon^{b}{_c} \right]=0 \ ,\ee
for arbitrary $\tau$'s. Think of the two--dimensional sphere as a
 submanifold of
 ${\Bbb R}^3$. By a rotation of the coordinate axes we can
 always achieve $F=\lambda \cos\theta$, for some constant
 $\lambda$. Equation \eq{muid} at a point $p_0$ lying on the
 equator, $p_0=(\theta=\pi/2,\phi_0)$, with $a=\theta$, $b=\theta$ reads
\be
\label{muid.1}
2\lambda {\cal D}_\phi \tau^{\phi\phi}=0 \ .  \ee Consider the
smooth traceless symmetric tensor
field $\tau_{ab}dx^a dx^b=\rho( (d\theta)^2 -\sin^{2}\theta (d\phi)^2)
+ 2\sigma d\theta\,d\phi $, with $\rho$ and $\sigma$ -- smooth
functions on $S^2$, supported near the equator, and satisfying
$\rho(p_0)=\sigma(p_0)=0$. Eq.\ \eq{muid.1} implies
$$
\lambda\partial_\phi\rho(p_0)=0\ ,
$$
for all such functions $\rho$, so clearly $\lambda = 0$, and 
we finally get
\begin{equation}
\label{E2.10}
 F = 0\ .
\end{equation}
Define
$$\Psi[f,\,x^a] = \hat{h}[f+\frac14{\cal D}_a{\cal D}_b
\chi^{ab},\,\chi_{ab}=0,\,x^a].$$
(\ref{E2.8}) and (\ref{E2.10}) give
$$\int_{S^2} \Psi[M-\frac14{\cal D}_a{\cal D}_b \chi^{ab}] = \int_{S^2}
\hat{h}[M,\,\chi_{ab}] = H[g]\ ,$$
which is what had to be established. \hfill $\Box$

\section{Super--translation invariance}
\label{BMSinvariance}
Theorem \ref{T.1} does not quite lead to the Trautman--Bondi mass as a
preferred quantity in the class of functionals considered in that
theorem, as it still contains an
arbitrary function $\Psi$ of $M-\frac{1}{4}\bh^{ac}\bh^{bd}{\cal
  D}_a{\cal D}_b \chi_{cd}$ and a finite number of its angular
derivatives. Let us show that the further requirement of {\em passive
  super--translation invariance\/} of $H$ can be used to obtain that
desired conclusion. Here the qualification ``passive'' refers to the
fact that we use a different Bondi coordinate system but we integrate
on the same cut of \Scri. More precisely,
consider a functional $H$ as in Theorem \ref{T.1}. We can calculate
the value of $H$ at a cross-section $S^2$ for a metric $g$, and
compare the result with $H$ calculated on the same cross-section of
\Scri\ for the same metric with a different Bondi parametrization,
differing by a (finite, or infinitesimal) BMS super-translation.  Let
${\cal S}$ denote a given cut of \Scri, which in some Bondi coordinate
system $(u,\theta,\phi)$ on \Scri\ is given by the equation $u=0$, and
set \be H({\cal S})=\int_{S^2} \Psi(M-\frac{1}{4}\bh^{ac}\bh^{bd}{\cal
  D}_a{\cal D}_b \chi_{cd})(u=0,\theta,\phi) \dx{}^2\mu\ .
\label{integral} \ee Consider another Bondi coordinate system $(\bar
u,\bar \theta,\bar \phi)=( u-\alpha(\theta,\phi),\theta,\phi)$, with
corresponding functions $\bar M$, $\bar\chi_{\bar a\bar b}$, {\em
  etc.\/} As shown in Appendix \ref{supertranslacje} (see
also \cite{JJAPP98}), we have 
 \be\label{trule} \overline{\left[4M - {
      \chi}^{ab}{_{||ab}}\right]} (u,\theta,\phi)= \left[4M - {
    \chi}^{ab}{_{||ab}}+
  {\dtwo}({\dtwo}+2)\alpha\right](u,\theta,\phi) \ .\ee
 The overbar in the left hand side of the last equation denotes 
the  quantity $4M -
 { \chi}^{ab}{_{||ab}}$ calculated in the barred Bondi frame, using
the barred Bondi functions $\bar M$, {\em etc.\/}
The requirement that $H({\cal S})$, calculated in the unbarred Bondi
coordinate system, coincides with $H({\cal S})$, as calculated in the
barred Bondi coordinate system,  gives thus the equation
\begin{equation}
\label{F.3}
\forall \alpha \quad \int_{S^2} \Psi[M-\frac14{\cal D}_a{\cal D}_b
\chi^{ab}]\dx{}^2\mu = \int_{S^2} \Psi[M-\frac14{\cal D}_a{\cal D}_b
\chi^{ab} + \frac14({\cal D}_a {\cal D}^a + 2){\cal D}_b{\cal
D}^b\alpha]\dx{}^2\mu.
\end{equation}
(It should be emphasized that ${\cal S}$ is {\em not}  given by
the equation $\bar u= 0$. We are {\em not} requiring that the value
$H(\bar{{\cal S}})$ 
of $H$, calculated on the cut $\bar{{\cal S}}=\{\bar u= 0\}$,
coincides with that  of $H({\cal S})$. That last condition would
be the requirement that the value of $H$ does not
depend on the cut under consideration, which is of course
absurd in the radiating regime.)
Now, 
elementary considerations
 using spherical harmonics show that $\chi=({\cal D}_a {\cal D}^a +
 2){\cal D}_b{\cal D}^b\alpha$ is an arbitrary function such that 
 $P\chi=\chi$, where $P$ is the projection operator introduced in
 equation (\ref{M.1}). If we replace $\alpha$ by $t\alpha$ in eq.\
\eq{F.3},  differentiate with respect to $t$, and set $t=0$, we obtain thus
$P \frac{\delta\Psi}{\delta f}=0$.  
It follows that there exist constants $w^\mu$, $\mu=0,1,2,3$ such that
${\delta\Psi\over\delta f}=(w^0 +w^k n_k) {/4\pi}$,
$n_k:=\frac{x_k}r$ being an 
orthogonal (but not orthonormal) basis in the space $SH^1$ of the
$\ell=1$ spherical harmonics. The condition that 
${\delta\Psi\over\delta f}$ be nonnegative gives
 $w^0 +w^k n_k \geq 0$ for all $n_k\in S^2$. That will hold if and
only if
$w^0 \geq |w|$,
 where $|w|=\sqrt{\delta^{kl}w_k w_l}$, so that one may think of
$w^\mu$ as of a 
future timelike vector.
We have thus obtained $\Psi(f)=(w^0 +w^k n_k) f {/4\pi}$, and finally
\be H=\frac{1}{4\pi}\int_{S^2} (w^0 +w^k n_k)(M-\frac 14
{\chi^{ab}}_{||ab})\, d^2\mu= 
  \frac{1}{4\pi}\int_{S^2} (w^0 +w^k n_k) M\,d^2\mu \ .
\label{F.4}
\ee
Equation \eq{F.4} has the clear interpretation that $H$ is the
Trautman--Bondi mass as measured with respect to a frame with time-like
four--velocity vector $(w^0,w^i)$, which can be checked from
the transformation properties of Bondi coordinate systems under
(passive) Lorentz transformations. For completeness we analyze that
question in Appendix \ref{boost}.  

The results of this section and Theorem \ref{T.1} imply the following:

\begin{Theorem}
\label{T.2}
Let $H$ be a functional of the form
\begin{equation}
\label{E3.1}
H[g,\,u] = \lim_{\begin{array}{c} {\scriptstyle \rho \rightarrow
\infty}\\[-5pt] {\scriptstyle t-\rho=u}
\end{array}}
\int_{S(t,\,\rho)} H^{\alpha\beta}(g_{\mu\nu},\,
g_{\mu\nu,\sigma},\,\ldots,\,g_{\mu\nu,\sigma_1\ldots\sigma_k}) \dx
S_{\alpha\beta},
\end{equation}
where the $H^{\alpha\beta}$ are twice differentiable functions of
their arguments. Suppose that $H$ is monotonic in $u$ for all vacuum
metrics $g_{\mu\nu}$ for which $H$ is finite, provided that
$g_{\mu\nu}$ satisfies
\begin{equation}
\begin{array}{c}
\displaystyle{g_{\mu\nu} = \eta_{\mu\nu} +
\frac{h^1_{\mu\nu}(u,\,\theta,\,\phi)}{r} +
\frac{h^2_{\mu\nu}(u,\,\theta,\,\phi)}{r^2} + o(r^{-2})}
\ ,\\[10pt]
\displaystyle{\partial_{\sigma_1} \ldots \partial_{\sigma_k}( g_{\mu\nu} -
\frac{h^1_{\mu\nu}(u,\,\theta,\,\phi)}{r} -
\frac{h^2_{\mu\nu}(u,\,\theta,\,\phi)}{r^2}) = o(r^{-2}),\quad
1\leq i \leq k,}
\end{array}
\end{equation}
for some $C^k$ functions 
$h^a_{\mu\nu}(u,\,\theta,\,\phi)$, $a=1,2$. 
If $H$ is invariant under passive BMS
super-translations, then the numerical 
value of $H$ equals (up to a proportionality constant) the Trautman--Bondi
mass.
\end{Theorem}

\noindent{\bf Proof:} If $H$ is monotonic for all such metrics, then
it is monotonic for Bondi-Sachs type metrics (\ref{G.2}) for which a
quasi-Minkowskian coordinate system (\ref{G.3}) has been introduced.
As discussed at the beginning of Section \ref{monotone}, for such
metrics (\ref{E3.1}) can be written as a quadratic polynomial in the
relevant fields, linear in $N^a$, so that Theorem
\ref{T.1} applies.  Now
the asymptotic behaviour of the functions appearing in the metric
\eq{G.2} shows that any quadratic terms in $M$ that could possibly
survive in the limit $r\to\infty$ come with no angular derivatives
acting on $M$.  The definiteness of the variational derivative of
$\Psi$, where $\Psi$ is given by Theorem \ref{T.1}, together with
Lemma \ref{L.1}, implies then
that $\Psi$ is necessarily linear, and
the result follows from the argument leading to (\ref{F.4}). \hfill $\Box$

Note that the trivial monotone functional, namely $H \equiv 0$, is
contained in the result above, the relevant constant of
proportionality being zero.

\section{General cuts of \Scri}
\label{generalcuts}

So far we have been considering the TB mass of those cuts of \Scri\
which are given by the equation
$u=0$. Consider
now a cut ${\cal S}$ of $\Scri$ which, in Bondi coordinates, is given
by the equation 
$$
{\cal S}=\{u=s(\theta,\phi)\}\ ,
$$ 
for some, say smooth, function $s$ on $S^2$. Theorem \ref{T.1},
together with the discussion of the previous section,
suggests that it is natural to define
\be  m_{TB}({\cal S}):= {1\over 16\pi}\int_{S^2}
\left(4M-{\chi}{^{ab}}{_{||ab}}\right)(u=s(\theta,\phi),\theta,\phi)\sin \theta
\rd\theta \rd\phi \label{dmTBm}
\ee
\be  p^k({\cal S}):= {1\over 16 \pi}\int_{S^2}
\left(4M-{\chi}{^{ab}}{_{||ab}}\right)(u=s(\theta,\phi),\theta,\phi){n^k}
\sin \theta \rd\theta \rd\phi \ ,\label{dpTBm}
\ee
where $n^k$, $k=1,2,3$ denotes the functions $\sin \theta \cos \phi$, 
$\sin \theta \sin \phi$ and $\cos \theta$, in that order. 
We have:
\begin{itemize}
\item As observed in Section \ref{BMSinvariance} ({\em cf.\/} eq.\
\eq{F.4}), equation \eq{dmTBm} reduces to the standard
Trautman--Bondi--Sachs definition when $s\equiv 0$.
\item It also follows from what is said in the previous section that
the quantities \eq{dmTBm}--\eq{dpTBm} are invariant under passive BMS
super--translations.
\item Equation \eq{F.4} together with passive super--translation
invariance and the discussion of Appendix \ref{boost} imply that the
quantities $(p^\mu)=(m_{TB},p^k)$ transform as a Lorentz vector under
those boosts which map ${\cal S}$ into itself.
\item 
The definitions (\ref{dmTBm})--(\ref{dpTBm}) allow us to define a flux
of energy--momentum through a subset of $\Scri^+$ bounded by two
cross-sections thereof. More precisely, let
${\cal S}_i$, $i=1,2$ be two cross-sections of $\Scri^+$ which are
graphs over the cut $u=0$:
$$
{\cal S}_i=\{u=s_i(\theta,\phi)\}\ ,
$$ 
and  let $N\subset \Scri^+$ be such that
$\partial N=s_2(S^2) \cup s_1(S^2)$. From the
definition (\ref{dmTB}) and the relation (\ref{G.6}) we have
\[
m_{TB}({\cal S}_2)-m_{TB}({\cal S}_1)=\frac1{16\pi}\int_{\partial N}
\left(4M-{\chi}{^{ab}}{_{||ab}}\right)\sin \theta
 \rd\theta \rd\phi =\]
\be =-\frac1{32\pi}\int_{N}
{\chi}_{ab,u}{{\chi}^{ab}}_{,u} 
\sin \theta \rd u \rd\theta \rd\phi\ ,
\label{fN} \ee
which can be thought of as a flux of energy through N. A similar
formula holds for the space--momentum $p^k$ defined by (\ref{dpTB}):
\[  p^k({\cal S}_2)-p^k({\cal S}_1)=\frac1{16\pi}\int_{\partial N}
\left(4M-{\chi}{^{ab}}{_{||ab}}\right)n^k
\sin \theta \rd\theta \rd\phi =\] 
\be =-\frac1{32\pi}\int_{N}
{\chi}_{ab,u}{{\chi}^{ab}}_{,u} n^k
\sin \theta \rd u \rd\theta \rd\phi\ .
\label{fN1}
\ee We note that the existence of a flux formula is a rather trivial
property, since one can always take the $u$ derivative of any
integrand to obtain a flux. The interest of the above formulae stems
from the fact that ${\chi}_{ab,u}$ is invariant under (passive)
super--translations, so that the fluxes \eq{fN}--\eq{fN1} also share
this property.
\item Passive super--translation invariance together with the flux
formulae \eq{fN}-\eq{fN1} imply that in a stationary space--time the
four--momentum $p^\mu$ defined by \eq{dmTBm}--\eq{dpTBm} is ${\cal S}$
independent. 
In particular $p^\mu$ vanishes in Minkowski space--time,
independently of the cut $\cal S$.  
\end{itemize}
\section{Polyhomogeneous metrics}\label{polhmg}

Having established the preferred role played by the Trautman--Bondi
mass, it is of interest to enquire under what weaker asymptotic
conditions one can still obtain a definition of mass which is finite
and monotonic in $u$. Recall that in \cite{ChMS} an {\em ad hoc\/}
definition of mass was given for all Bondi-type metrics with a
``polyhomogeneous \Scri'', and that mass was shown there to be
monotonic.  Similarly it was checked in \cite{LauYorkBrown} that for a
class of asymptotically flat asymptotically vacuum
space-times\footnote{The class of metric considered in
  \cite{LauYorkBrown} includes the metrics polyhomogeneous of order
  $2$ (see \cite{ChMS} and below for definitions).} the energy
expression defined in \cite{YorkBrown} converges to an appropriately
defined Bondi mass.  {}From a field theoretic point of view it is
natural to define mass in terms of an integral, as considered in
Theorem \ref{T.2}, using {\em e.g.} the Freud potential, where the
$H^{\alpha\beta}$ of equation (\ref{G.1}) is given by the
expression ({\em cf.\ e.g.\/} \cite{Trautman:Witten})
\begin{eqnarray}
  &  \label{Freud2}
 H^{\mu\nu} = {\mbox{{\eufm U}}^{\mu\nu}}_\alpha X^{\alpha}
 \ ,& \\ & {\mbox{{\eufm U}}^{\mu\nu}}_\alpha = \displaystyle{\frac{1}{ 16\pi\sqrt{|\det  g_{\rho\sigma}|}}} g_{\alpha\beta}(|\det g_{\rho\sigma}|
   g^{\beta[\mu}g^{\nu]\lambda})_{,\lambda}\ ,&\label{Freud2.0}
\end{eqnarray}
with $X^\mu = \delta^\mu_0$.
Inserting the metric \eq{PE.3} into \eq{Freud2.0}, with 
$X^\alpha \partial_\alpha = \partial _u$ and with $h_{ab}$
parametrized as in \eq{PE.4}, one obtains via a {\sc Sheep}
\cite{MSMM} calculation
\begin{eqnarray}
\nonumber
\int_{u=u_0,r=r_0}{\mbox{{\eufm U}}^{\mu\nu}}_\alpha
X^{\alpha}dS_{\mu\nu}&  = & {1\over 16 \pi}\int_{S^2}
\Big\{-2 V +2re^{2\beta} \cosh(2\gamma)\cosh(2\delta)
\\ \nonumber & & 
-r^4 e^{-2\beta}\Big[ {\partial U^\theta \over
    \partial r} \big(U^\theta e^{2\gamma}\cosh(2\delta)+U^\phi
    \sinh(2\delta) \sin (\theta)\big) 
\\ \nonumber
& & 
+ \sin (\theta) {\partial U^\phi \over
    \partial r} \big(U^\phi e^{-2\gamma}\cosh(2\delta)\sin (\theta)+U^\theta
    \sinh(2\delta)\big) 
\Big]
\\  & & + r^2 {\cal D}_aU^a
\Big\}\Big|_{u=u_0,r=r_0} \sin (\theta) d\theta \, d\phi 
 \label{reducedenergy0}
\\ \nonumber & = &  {1\over 16 \pi}\int_{S^2}
\Big\{2 (r-V) 
\\ \nonumber & &  +2r(1-e^{2\beta} \cosh(2\gamma)\cosh(2\delta))
-r^4 e^{-2\beta}h_{ab} {\partial U^a \over
    \partial r} U^b
\\  & & + r^2 {\cal D}_aU^a
\Big\}\Big|_{u=u_0,r=r_0} \sin (\theta) d\theta \, d\phi \ .
 \label{reducedenergy}
\end{eqnarray}
More precisely, this formula is obtained by ``covariantizing'' (as
described in \cite{ChAIHP}) 
eq.\ \eq{Freud2.0} with the following  flat background metric $\eta$:
\[
\eta_{\mu\nu}dx^\mu dx^\nu = -du^2 - 2du\, dr + r^2(d\theta^2 +\sin^2(
\theta)\, d\phi^2)\ .
\]
Eq.\ \eq{reducedenergy} is exact; no hypotheses about the asymptotic
behaviour of the quantities involved have been made. Note that the
last term in eq.\ \eq{reducedenergy} integrates out to zero.
We shall say that a metric is {\em polyhomogeneous of order $k$} if in
the Bondi coordinates (\ref{G.2}) the functions $h_{ab}$ have a
polyhomogeneous expansion (see \cite{ChMS} for definitions) in which
the $\ln r$ terms start at a power $r^{-k}$:
$$h_{ab} = \bh_{ab} + \frac{h^1_{ab}}{r}+\ldots
+\frac{h^{k,n}_{ab}\ln^n r}{r^k}+\frac{h^{k,(n-1)}_{ab}\ln^{(n-1)}
  r}{r^k}+ \ldots$$ 
Consider first  metrics which are
polyhomogeneous of order $2$. We have then $\gamma=O(r^{-1})$,
$\delta=O(r^{-1})$ and it follows from the Einstein equations as
written out {\em e.g.\/} in \cite[Appendix C]{ChMS}\footnote{There are
  unfortunately some misprints in Appendix C of \cite{ChMS}: 1) The last
  term in eq. (C4), ${1\over 2} r^2 \cosec \theta (W_{13} + 4
  W_3)$, should be replaced by ${1\over 2} r \cosec \theta (rW_{13} + 4
  W_3)$; 2) in the 8th line of Eq.\ (C6) the factor $4$ in front of
  the term $4 \gamma_1 \gamma_2  U$ should be replaced by $2$.}
that $\beta=O(r^{-2})$, $U^a=O(r^{-2})$, $\partial U^a/\partial
r=O(r^{-3})$  and $r-V =O(1)$. Eqs. \eq{G.1} and 
\eq{reducedenergy} then give
\be
H[u_0,g] = \lim_{r \to \infty}{1\over 8
  \pi}\int_{S^2}(r-V)\sin(\theta)\, d\theta d\phi\ ,
\label{finalenergy}
\ee 
which is the standard Bondi integral.
Consider, next, metrics which are polyhomogeneous of order $1$. In
that case one has $\gamma=O(r^{-1}\ln^N r)$, $\delta=O(r^{-1}\ln^N r)$
for some $N$. The Einstein equations imply (see the proof of Prop.\ 
2.1 in \cite{ChMS}) that $\beta=O(r^{-2}\ln^{2N} r)$,
$U^a=O(r^{-2}\ln^{N} r)$, $\partial U^a/\partial r=O(r^{-3}\ln^{N} r)$
and $r-V =O(\ln^{N_V} r)$ for some $N_V$.
Eqs. \eq{G.1} and \eq{reducedenergy} lead
again to \eq{finalenergy}. At first sight it appears that the integral
at the right hand side of \eq{finalenergy} might diverge for some
vacuum metrics which are polyhomogeneous of order $1$. However,
careful study of the leading terms in the Einstein equations shows
that those terms in $V$ which are linear combinations of $\ln^i r$ are
divergences, so that their integral over a sphere vanishes.
Thus
the Freud integral always converges to the monotonic mass expression
considered in \cite{ChMS}.  Remarkably, the polyhomogeneous case of
order $k \geq 1$ always has a finite energy.

Let us mention that for metrics which are polyhomogeneous of order
$k\ge 2$ the Freud integral can be given a Hamiltonian interpretation
--- this will be discussed elsewhere.

\section{Closing remarks}\label{closing}

We have shown that every functional of the
fields which is monotonic in time in a certain class of functionals
for all metrics ``having a piece of $\Scri$'' is proportional to the
Trautman--Bondi mass. The key ingredient of our proof was the Friedrich--Kannar
construction of space--times ``having a piece of $\Scri$''. Now in
general the space--times we have constructed in the proof above will
not have any reasonable global properties. For example, in Lemma
\ref{L.1} the function $M$ could be chosen to be negative. In such a
case one expects, from the positive TB mass conjecture, that
the space--time constructed in Lemma \ref{L.1} will have no extension
with complete Cauchy surfaces. Now the property of having such Cauchy
surfaces is a starting point of any standard Hamiltonian analysis, and for this
reason it would be rather useful to have an equivalent of Lemma
\ref{L.1} in which well behaved space--times are constructed. We
expect that a result of that kind can be proved, under some mild (yet
to be determined) restrictions on the function $M$ (such as {\em
e.g.\/} positivity), and we are planning to investigate this problem
in the future. 

Let us finally mention that using similar ideas to those presented
here one can prove
related results for other field theories, such as {\em e.g.\/} Maxwell
theory, or for scalar fields. More precisely, for a scalar field one
has the following:

\begin{Theorem}
  \label{Tscalar}
  The only functional $F$, in the class of functionals defined in the
  Introduction, of a scalar field $\phi$ on Minkowski space--time,
  which is monotonic in retarded time for all solutions of the
  massless linear wave equation, and which is a Hamiltonian for
  the dynamics on a hyperboloid $\Sigma$, is the integral $H$ of the
  standard energy--momentum tensor over $\Sigma$.
\end{Theorem}

To prove this one uses an equivalent of Lemma \ref{L.1} which, for a
scalar field on Minkowski space--time, can be easily modified to
obtain globally defined solutions. The question of how to define a
symplectic structure for dynamics on hyperboloids will be discussed
elsewhere \cite{CJK}. The requirement that the functional considered
is a Hamiltonian leads to the conclusion that $F$ differs from $H$ by
a boundary integral. Using arguments similar to the ones presented in
this paper (and actually rather simpler, as the corresponding
equations on \Scri\ are much simpler in the case of a scalar field)
one then proves \cite{CJMPRL} that all the boundary integrands, in the
case of the scalar field, which have the right monotonicity
properties, have to integrate out to zero. Minkowski space--time above
can be replaced by any Lorentzian manifold which has sufficiently
regular conformal completions.

Let us finally mention that one can set up a Hamiltonian framework in
which some of the problems related to the Ashtekar--Streubel or
Ashtekar--Bombelli--Reula approaches, listed in section
\ref{nonunique}, are avoided \cite{CJK}. Unsurprisingly, the
Hamiltonians one obtains in such a formalism are again not unique, but
the non--uniqueness can be controlled in a very precise way. The
Trautman--Bondi mass turns out to be a Hamiltonian, and
an appropriate version of the uniqueness Theorem \ref{T.1} proved
above can be used to single out the TB mass amongst the family of all
possible Hamiltonians.

{\bf Acknowledgements} PTC is grateful to the E.~Schr\"odinger
Institute in Vienna for hospitality and financial support during part
of work on this paper. JJ wishes to thank the R\'egion Centre for
financial support, and the Department of Mathematics of the Tours
University for hospitality during work on this paper. We acknowledge a
collaboration with L.~Andersson at an early stage of work on the
questions raised here. We are grateful to P.~Tod for bibliographical
advice.

\appendix
\section{Trautman's definition of mass in the radiation regime}
\label{Trautman}

In \cite{T}\footnote{The first chapter of \cite{Tlectures} is a
  slightly expanded version of \cite{T}.} Trautman considers
  gravitational fields for which a coordinate system exists in which
  the metric can be written in the form 
  \begin{eqnarray}
    \label{9}
    & g_{\mu\nu} = \eta_{\mu\nu}+O(r^{-1}), \qquad g_{\mu\nu,\rho}=
    h_{\mu\nu} k_\rho+O(r^{-2})\ ,& \\ \label{10} &
    (h_{\mu\nu}-\frac{1}{2}\eta_{\mu\nu} \eta^{\rho\sigma}
    h_{\rho\sigma})k^\nu = O(r^{-2})\ .&
  \end{eqnarray}
  Here the functions $h_{\mu\nu}$ satisfy $h_{\mu\nu}=O(r^{-1})$,
  while the null vector field $k_\nu$ is defined as follows: Let
  $\sigma $ be a spacelike hypersurface, and define $n^\mu$ to be a
  unit space--like vector lying in $\sigma$ perpendicular to the
  sphere $r=\mbox{const}$, and pointing outside it. Trautman
  defines $k^\nu$ to be $n^\nu+t^\nu$, where $t$ denotes a unit
  time--like vector normal to $\sigma$, such that $t^0>0$.

Trautman shows that under the conditions \eq{9}--\eq{10} the integral
at the right--hand--side of the equation
\begin{equation}
  \label{5}
  P_\mu[\sigma]=\oint_S {\mbox{{\eufm U}}^{\nu\lambda}}_\mu dS_{\nu\lambda}
\end{equation}
exists\footnote{\label{footnotenodef} It is
clear that $S$ in \eq{5} is understood as ``a boundary of $\sigma$ at
infinity'', defined as far as integration  is concerned by a
limiting process. In the section in which
  he talks about radiating fields Trautman does not give a precise
  definition of what $S$ is.} and is finite because of cancellations among the
divergent terms. Here ${\mbox{\eufm U}^{\nu\lambda}}_\mu$ is
the Freud potential given in Eq.\ \eq{Freud2.0}.  Next, Trautman shows
that $P_\mu[\sigma]$ is coordinate independent in the following sense:
Let a new coordinate system ${x^\prime}^\nu $ be given by the
equations 
\begin{equation}
  \label{11}
  x^\nu \to {x^\prime}^\nu = x^\nu+ a^\nu\ ,
\end{equation}
with $a^\nu$ satisfying
\begin{equation}
  \label{12}
 a^\nu = o(r)\ , \qquad a_{\nu,\mu}=b_\nu k_\mu + O(r^{-2})\ , 
\end{equation}
where
$$ a_\nu= \eta_{\nu\mu}a^\mu\ , \qquad b_\nu = O(r^{-1})\ ,$$ and
\begin{equation}
  \label{13}
  a_{\nu,\mu\rho}=b_{\nu,\mu} k_\rho + O(r^{-2})\ ,\qquad b_{\nu,\rho}
  = O(r^{-1})\ .
\end{equation}
Those coordinate transformations preserve the boundary conditions
introduced above. Trautman notices that under those transformations
the integrand in \eq{5} changes by terms which are $O(r^{-3})$, so that
$P_\mu[\sigma]$ itself remains unchanged.

In Section 4 of \cite{T} Trautman gives the formula for the total
energy and momentum, which he calls $p_\mu$, radiated between two
hypersurfaces $\sigma$ and $\sigma^\prime$,
\begin{equation}
  \label{15.0}
  p_\mu= P_\mu[\sigma] - P_\mu[\sigma^\prime]=\int_\Sigma
  {{\mbox{\eufm t}}_\mu}^\nu dS_\nu\ , 
\end{equation}
under the hypothesis that the energy--momentum tensor of matter fields
gives no contribution on $\Sigma$. Here
\begin{equation}
  \label{16}
  {\mbox{\eufm t}_\mu}^\nu = \tau k_\mu k^\nu + O(r^{-3})\ ,
\end{equation}
where 
\begin{equation}
  \label{17}
  4 \kappa \tau = h^{\mu\nu}(h_{\mu\nu}-{1\over 2}
  \eta_{\mu\nu} \eta^{\rho\sigma}h_{\rho\sigma})\ ,
\end{equation}
and $\kappa$ is the constant of proportionality between the Einstein
tensor and the energy--momentum tensor, and it is clear that the integral over
$\Sigma$ in Eq.\ \eq{15.0} is  defined by a limiting
process \footnote{In the section in which he talks about radiating
  fields Trautman does not give a precise definition of what $\Sigma$
  is. In a preceding section of \cite{T} where boundary conditions
  appropriate for spatial infinity are considered he uses the same
  equation to show that $P_\mu$ is conserved, and in that case he
  defines $\Sigma$ as ``a {time--like} ``cylindrical''
  hypersurface at spatial infinity''.}. He emphasizes that $\tau$ is
invariant with respect to the transformations \eq{11} and is {\em
  non--negative} by virtue of \eq{10}, so that $p_0\ge 0$. 

For our purposes we need to change the definition of $k_\mu$ given
above: we require $k_\mu$ to be a null vector field satisfying
\begin{enumerate}
\item $k_\mu$ is normal to the spheres $r=\{\mbox{const}\}$, future
  pointing and outwards--pointing;
\item $k_\mu$ satisfies the following asymptotic conditions:
$$ k^0 - 1 = O(r^{-1})\ ,\qquad k^i-{x^i\over r} = O(r^{-1})\ .$$
\end{enumerate}
(This is compatible with Trautman's definition if one takes $\sigma$
to be the hypersurface $\{x^0=\mbox{const}\}$ in the coordinate system
in which \eq{9}--\eq{10} hold. However, the hypersurfaces we consider here are
\emph{not} of this form.)  With this modification Eq.\ 
\eq{15.0}--\eq{16} together with positivity of $\tau$ are the
fundamental statement that on hypersurfaces which, in modern
terminology, ``intersect $\Scri^+$'' the energy can only be radiated away.
It should be emphasized that this is a more general statement than
that discussed by Bondi {\em et al.}\ and by Sachs four years later
\cite{BBM,Sachs}, as the boundary conditions \eq{9}--\eq{10} are {\em
  weaker} than those of \cite{BBM,Sachs}. Indeed, consider a
Bondi--Sachs type metric \eq{G.2}, with all the functions appearing
there satisfying the fall--off requirements of \cite{BBM,Sachs}. If
quasi--Minkowskian coordinates are introduced via the equations
\eq{G.3}, one finds that Trautman's conditions \eq{9}--\eq{10} hold
with $k_\mu = u_{,\mu}$. If $\sigma$ is taken to be the null
hypersurface $\{u=u_0\}$ (note that with our minor modification of the
definition of what $k_\mu$ is, the hypothesis that $\sigma$ is
spacelike is not needed any more in the above formalism) the
four--momentum $P_\mu[\sigma]$ defined by Eq.\ \eq{5} gives the Bondi
mass as defined in \cite{BBM,Sachs}.  If $\sigma^\prime$ is taken to
be another such null hypersurface, Eq.\ \eq{15.0} yields the Bondi
mass loss formula (integrated in $u$). Further, the coordinate
transformations \eq{11} comprise the BMS ``super-translations'': a
super--translation given by Eqs.\ \eq{aseq.1}--\eq{aseq.4} below
corresponds to a transformation \eq{11} with
$a^\mu=\phi^\mu(\theta,\phi) + O(1/r)$, for some appropriate functions
$\phi^\mu(\theta,\phi)$, so that $b_\mu$ in \eq{12} vanishes.

It should be pointed out that, as discussed in Section
\ref{polhmg} above, the fall--off conditions \eq{9}--\eq{10}
allow for a large class of metrics with polyhomogeneous asymptotics.
Last but not least, using the framework of \cite{T} reduces the
computational complexity of the proof of positivity of mass--loss, as
compared to several other frameworks, \emph{e.g.} the Bondi--Sachs
one.

\section{On some operators on $S^2$} 
\label{surjectivity}
 Let us denote by
$\dtwo$ the Laplace--Beltrami operator associated with the standard
metric on $S^2$, $\dtwo={\cal D}^a{\cal D}_a$. Let $SH^l$ denote the space of
spherical harmonics of degree $l$ ($g\in SH^l \Longleftrightarrow
 {\dtwo}g= -l(l+1)g$).
Consider the following sequence 
\[ 
\begin{array}{ccccccccc}
V^0\oplus V^0 & \stackrel{i_{01}}{\longrightarrow} & V^1
& \stackrel{i_{12}}{\longrightarrow} & V^2 
& \stackrel{i_{21}}{\longrightarrow} & V^1
& \stackrel{i_{10}}{\longrightarrow} & V^0 \oplus V^0 \ .
\end{array}
\]
Here $V^0$ is the space of, say, smooth functions on $S^2$,
 $V^1$ -- that of smooth covectors on $S^2$, and 
$V^2$ -- that of symmetric traceless tensors on $S^2$.
The various mappings above are defined as follows:
\[ i_{01}(f,g)=f_{||a}+\varepsilon_a{^b}g_{||b} \ ,\]
\[ i_{12}(v)= v_{a||b}+ v_{b||a}-\bh_{ab}v^c_{||c} \ ,\]
\[ i_{21}(\chi)= \chi_a{^b}{_{||b}} \ ,\]
\[ i_{10}(v)=\left( {v^a}_{||a}, \varepsilon^{ab}v_{a||b} \right) \ ,\]
where $||$ is used to denote the covariant derivative with respect to
the Levi--Civita connection of the standard metric $\bh_{ab}$ on
$S^2$.  The following equality holds
\be
\label{compid}
 i_{10} \circ i_{21} \circ i_{12} \circ i_{01} = ({\dtwo }({\dtwo
}+2))\oplus ({\dtwo }({\dtwo
}+2)) \ .\ee
Note that we have $i_{10}\circ i_{21}(\chi)=( {\chi^{ab}}_{||ab},
{{\varepsilon^{bc}}}{{\chi^{a}}_{b}}_{||ac})$ 
Consider the space
$\overline V^0:=[SH^0\oplus SH^1 ]^\perp $, where $\perp$ denotes $L^2$
orthogonality in $L^2(S^2)\cap C^\infty(S^2)$. Now the operator
${\dtwo }({\dtwo }+2)$ is surjective from $V^0$ to $\overline V^0$, so
that for any $\lambda\in \overline V^0$ there exists $f\in V^0$ such
that ${\dtwo }({\dtwo }+2)f = \lambda$. Consider the tensor field
$\chi= i_{12}\circ i_{01}((f,0))$, then \eq{compid} shows that
${\chi^{ab}}_{||ab}=\lambda$, which establishes surjectivity of the
double divergence operator, from the space of symmetric traceless
tensors to that of functions on the sphere which have no zero and
first harmonics. Similarly the tensor field $\chi= i_{12}\circ
i_{01}((0,g))$ shows that the map $V^2\ni \chi_{ab}\to
{{\varepsilon^{c}}_b}{\chi^{ab}}_{||ac}\in [SH^0\oplus SH^1 ]^\perp $
is surjective.  

To justify our claim, that the vector field $\hat \alpha$ given by
eq.\ \eq{E2.5} is the unique solution of eq.\ \eq{M.1.1}, consider the
sequence
\[ 
\begin{array}{ccccc}
V^0\oplus V^0 & \stackrel{i_{01}}{\longrightarrow} & V^1
& \stackrel{i_{10}}{\longrightarrow} & V^0 \oplus V^0 \ .
\end{array}
\]
It is easy to check that
\[ i_{10}  \circ i_{01} = {\dtwo } \oplus {\dtwo }\ , \]
so if
$\alpha_{a||b}\varepsilon^{ab},\; \alpha^a{_{||a}} \in (SH^0)^\perp$
 then there
exist $f,\, g \in (SH^0)^\perp$ such that $i_{01}(f,g)=\alpha$, and they are
the unique solutions in $(SH^0)^\perp$ of the equations:
\[ {\dtwo }f=\alpha^a{_{||a}} , \quad 
{\dtwo }g=\alpha_{a||b}\varepsilon^{ab} \ .\]
Our claim follows immediately from this observation.
\label{AppA}

\section{Super-translations}\label{supertranslacje}
As in Appendix \ref{AppA} we use the notation 
$   f_{||a} \equiv{\cal D}_a f  $,  
$ {\dtwo}\equiv{\cal D}^a {\cal D}_a$.

Consider a super-translation which in an appropriate coordinate system
on  \Scri\ reduces to a
transformation
$u\to u-\alpha(\theta,\phi)$, for some, say smooth, function $\alpha$
on $S^2$, with the angular coordinate being left invariant.
The super-translation can be extended from \Scri\ to a neighborhood
thereof in the physical space--time using
 Bondi coordinates ({\em cf.\/} eq.\ (\ref{G.2})). This  leads to the
following  asymptotic expansions (see also \cite[p. 119]{Sachs}): 
\be
{\overline x}^a=x^a +\frac 1r \alpha^{||a} -\frac 1{2r^2} \left( {
 \chi}^{ab}\alpha_{||b} -2\alpha^{||ab}\alpha_{||b}+\breve{\Gamma}^a{_{bc}}
\alpha^{||b}\alpha^{||c}\right) +\ldots \ ,\label{aseq.1}\ee
\be {\overline u}= u-\alpha -\frac 1{2r}\alpha^{||a}\alpha_{||a}
 + \frac 1{4r^2} \left[ { \chi}^{ab}\alpha_{||a}\alpha_{||b} -
\alpha^{||a}\left(\alpha_{||b}
\alpha^{||b}\right)_{||a}\right]+\ldots\ ,
\label{aseq.2}\ee
\[ {\overline r}=r-\frac 12 {\dtwo}\alpha + \frac 1{2r}
\left[ {{ \chi}^{ab}}_{||b}\alpha_{||a}+\frac12 {
    \chi}^{ab}\alpha_{||ab} +\frac12 {
    \chi}^{ab}{_{,u}}\alpha_{||a}\alpha_{||b} -\frac 12 \alpha^{||ab}
  \alpha_{||ab} - \alpha_{||a} \alpha^{||a}+ \right. \] \be \left.
  +\frac 14 ({\dtwo}\alpha)^2 - \left( {\dtwo}\alpha\right)^{||a}\alpha_{||a} \right]+\ldots \ ,
\label{aseq.4}
\ee
where $\breve{\Gamma}^a{_{bc}}$ is the connection defined by the
metric $\bh_{ab}$. From
those formulae we obtain the transformation laws for $ \chi$ and $M$:
\[ {\overline M}(\bar{u} =
 u-\alpha(x^a),\,x^a) = \left[ M + \frac 12 { \chi^{ab}_{,u}}_{||b}\alpha_{||a}+\frac14 \chi^{ab}_{,u}\alpha_{||ab} + \frac14{ \chi}^{ab}_{,uu}
\alpha_{||a} \alpha_{||b}\right](u,\,x^a)\ ,\]
\[ \overline{ \chi}_{ab}(\bar{u} =
 u-\alpha(x^c),\,x^c)=\left[ { \chi}_{ab} - 2\alpha_{||ab}+
{ \bh}_{ab}{\dtwo}\alpha \right](u,\,x^c) \ .\] 
Consider the quantity $ \overline{{ \chi}}^{ab}{_{||\bar a\bar b}}$, where 
${||\bar a\bar b}$ denotes covariant derivatives with respect to the
transformed coordinates,
$\partial_{\bar a}={\partial_a}+\alpha_{,a}{\partial_0}$. Note
that the occurence of $u$ derivatives in $\partial_{\bar a}$ will
introduce $u$ derivatives of $\chi_{ab}$ in the transformation formula
for this quantity, and one finds that 
the  combination $4M - { \chi}^{ab}{_{||ab}}$ has a
simple transformation law with respect to the super-translations:
\[ \overline{\left[4M -
 { \chi}^{ab}{_{||ab}}\right]} (\bar{u} =
 u-\alpha(\theta,\phi),\,\theta,\,\phi)= \left[4M - {
\chi}^{ab}{_{||ab}}+ {\dtwo}({\dtwo}+2)\alpha\right](u,\theta,\phi) \
.\] The overbar in the left hand side of the last equation denotes the
corresponding quantity calculated in the new Bondi frame.  Note that
while the equations \eq{aseq.1}--\eq{aseq.4} had only an asymptotic
character in $1/r$, the last three equations are exact; in particular
no smallness conditions on $\alpha$ have been imposed.

\section{Boost--transformations and $p^\mu$}\label{boost}
Let $\Lambda$ be  a boost-transformation
 with boost parameter $\nu$; by an appropriate choice of space--cordinates we can choose it to act along the $z$ axis. In coordinates \eq{G.3}
on Minkowski space--time one has
\[ \overline u =\frac{u}{\cosh\nu-\sinh\nu\cos\theta} +O({u\over r})\, ,
\quad \tan\frac{\overline\theta}2 =\E\nu \tan\frac\theta 2+O({u\over
r})\ , \]
with $\phi$ remaining unchanged.
It follows that 
on  $\Scri$ the boost $\Lambda$ reduces to the transformation
\be
\label{BC.0} \overline u =\frac{u}{\cosh\nu-\sinh\nu\cos\theta} \, ,
\quad \tan\frac{\overline\theta}2 =\E\nu \tan\frac\theta 2 \, ,
\quad \overline \phi= \phi\ . \ee
It is natural to interpret \eq{BC.0} as the definition of the action
of the Lorentz boost $\Lambda$ on \Scri\ for general space--times
admitting a \Scri.

Equation \eq{BC.0} leads to the following transformation laws
\[ \partial_{\overline u}= (\cosh\nu-\sinh\nu\cos\theta)\partial_u \ , \]
\[ \partial_{\overline\theta}=u \sinh\nu \sin\theta \partial_u +
(\cosh\nu-\sinh\nu\cos\theta)\partial_\theta \ ,\]
\be \label{BC.1} \sin\overline\theta
=\frac{\sin\theta}{\cosh\nu-\sinh\nu\cos\theta} \ , \qquad
\rd\overline\theta = \frac{\rd\theta}{\cosh\nu-\sinh\nu\cos\theta}\ , \ee 
\[ \cos\overline\theta =
\frac{\cosh\nu\cos\theta-\sinh\nu}{\cosh\nu-\sinh\nu\cos\theta}\ . \] 
From \eq{BC.1} one obtains the well known statement, that boosts
induce conformal transformations of ``spheres at infinity'': if we
denote by $\psi$ the transformation which takes $(\theta,\phi)$ to
$(\overline\theta,\overline\phi)$, then
\be\label{BC.2}
\psi^*\bh_{ab}=\varphi^{-2}\bh_{ab}\ ,
\ee
with 
$$\varphi=\cosh\nu-\sinh\nu\cos\theta\ .
$$
We note that $\varphi$ is a linear combination of $\ell=0$ and $\ell=1$
 spherical harmonics. Set
\be
\label{BC.3}
\overline r = \varphi r\ .
\ee
The coordinate transformation \eq{BC.0}, \eq{BC.3} preserves the
 leading order behaviour of all the components of the metric \eq{G.2}. It
 follows from \cite{Tamburino:Winicour} (compare also \cite{ChMS}) that
 \eq{BC.0}, \eq{BC.3} can be extended to a neighbourhood of \Scri\
 while preserving the Bondi form of the metric \eq{G.2},  the
  hypersurface $u$=0 being mapped into the hypersurface $\overline u=0$.
From \eq{BC.0}, \eq{BC.3} and \eq{G.2} at $u=0$ one immediately
obtains 
\be
\label{BC.00}
\overline{M }=\varphi^3M\ ,
\ee so that
\[ \int_{S^2}\overline{M }\sin\overline\theta\rd\overline\theta\rd
\overline \phi =\int_{S^2}\
  M (\cosh\nu-\sinh\nu\cos\theta) \sin\theta\rd\theta\rd\phi\ .  \] It
  follows that the knowledge of the $\ell=0$ harmonics of $M$ is not
  sufficient to determine the $\ell=0$ harmonics of $\overline M$.  Let
  us set
\be  m_{TB}|_{u=0}:= {1\over4\pi}\int_{S^2}
M|_{u=0}\sin \theta
\rd\theta \rd\phi \ ,\label{dmTB}
\ee
\be  p^k|_{u=0}:= {1\over4\pi}\int_{S^2}
M|_{u=0}{n^k}
\sin \theta \rd\theta \rd\phi \ ,\label{dpTB}
\ee
where $n^k$, $k=1,2,3$ denotes the functions $\sin \theta \cos \phi$, 
$\sin \theta \sin \phi$ and $\cos \theta$, in that order.
Equations \eq{BC.0}, \eq{BC.3} and \eq{BC.00}  also yield
 \[ \overline{M} \cos\overline\theta
 \sin\overline\theta\rd\overline\theta =
  M  (\cosh\nu\cos\theta-\sinh\nu)
\sin\theta\rd\theta \ .\]   
Consequently we obtain the transformation  law 
 \be
\label{trm}
\overline m_{TB} = m_{TB}\cosh\nu - p^z \sinh\nu \ ,\ee
\be
\label{trz}
 \overline p^z = p^z\cosh\nu  - m_{TB}\sinh\nu \ . 
\ee
As the choice of the axis along which $\Lambda$ acts was arbitrary,
the set of numbers $(p^\mu)=(m_{TB},p^k)$ transforms as a
(contravariant) four--vector under the {\em passive} action of the
Lorentz group on
\Scri. It is therefore natural to  interprete $m_{TB}$ as 
the time component, and the $p^k$'s as space--components of an
energy--momentum four--vector $p^\mu$. We use the qualification
``passive'' above to emphasize the fact that such a simple transformation
property holds only for those Lorentz transformations which map a
chosen cross--section of \Scri\ into itself.

\section{Changes of the Noether charge induced by changes of the
Lagrange function} 
\label{transformation}
In this Appendix we wish to derive the transformation rule of the 
``Noether charge'' \eq{E.4}, when the Lagrange function is changed by
the addition of a term of the form \eq{Yf},
\begin{equation}
{\cal L} \longrightarrow \hat{\cal L}={\cal L} +R, \qquad \qquad
R\equiv \partial_\lambda Y^\lambda\ ,
\label{VAR.0}
\end{equation}
with $Y^\mu$ being a smooth function of the fields and their
derivatives up to order $k-1$. Letting $\Omega$ be an arbitrary domain of
$\R^n$ with smooth boundary and compact closure, we have
$$
\int_{\partial \Omega} Y^\mu dS_\mu = \int_{ \Omega}R\  d^nx\ .
$$
Integration by parts gives
\begin{eqnarray}
\int_{\partial \Omega}\sum_{i=0}^{k-1}\left(\frac{\partial Y^\mu}{\partial
\phi^A_{\alpha_1\ldots \alpha_i}} \right. &-&
\left. \sum_{j=0}^{k-i-1}(-1)^j\partial_{\beta_1}\ldots \partial _{\beta_j}
\left(
\frac{\partial R}{\partial
\phi^A_{\mu\alpha_1\ldots \alpha_i\beta_1\ldots \beta_j}}\right)\right)
\delta \phi^A_{\alpha_1\ldots \alpha_i} dS_\mu \nonumber \\
 &=&
\int_{ \Omega}\frac{\delta R}{\delta \phi^A} \delta \phi^A d^nx \ ,
\label{VAR.1}
\end{eqnarray}
where ${\delta R}/{\delta \phi^A}$ is the variational derivative
of $R$, for any smooth fields $\delta\phi^A$.  Eq.\ \eq{VAR.1} still
holds with $\Omega=\R^n$ if the $\delta\phi^A$'s are compactly
supported. In that case arbitrariness of the $\delta\phi^A$'s implies
$$
\frac{\delta R}{\delta \phi^A}= 0\ ,$$
which expresses the well known fact that the field equations are
unchanged by the above transformation of the Lagrange function. It
follows that
\be
\int_{\partial \Omega}\sum_{i=0}^{k-1}\left(\frac{\partial Y^\mu}{\partial
\phi^A_{\alpha_1\ldots \alpha_i}} -
\sum_{j=0}^{k-i-1}(-1)^j\partial_{\beta_1}\ldots \partial _{\beta_j}
\left(\frac{\partial R}{\partial
\phi^A_{\mu\alpha_1\ldots \alpha_i\beta_1\ldots \beta_j}}\right)\right)
\delta \phi^A_{\alpha_1\ldots \alpha_i} dS_\mu = 0\ .
\label{VAR.2}
\ee
It is convenient to choose a coordinate system $(x^\mu)=(x^1,v^a)$
such that $\partial \Omega$ is given by the equation $x^1=0$, the
$v^a$'s, $a=1,\ldots,n-1$ being coordinates     on $\partial
\Omega$. Define
$$ \phi^{A,m}_{a_1\ldots a_\ell}= \phi^{A}_{\scriptsize{\underbrace{
1\ldots 1}_{m \ \mbox{\scriptsize times}}}a_1\ldots a_\ell} \ ,
$$
$$
R_{A,m}^{a_1\ldots a_\ell} = 
\sum_{j=0}^{k-\ell-m}(-1)^j\partial_{\beta_1}\ldots \partial _{\beta_j}
\left(\frac{\partial R}{\partial
\phi^{A,m}_{a_1\ldots a_\ell\beta_1\ldots \beta_j}}\right) \ .
$$
Integration by parts in \eq{VAR.2} yields
\be
\int_{\partial \Omega}
\sum_{m=0}^{k-1}\sum_{i=0}^{k-m-1}(-1)^i\partial_{a_1}\ldots \partial_{a_i}\left(\frac{\partial Y^1}{\partial
\phi^{A,m}_{a_1\ldots a_i}} -R_{A,m+1}^{a_1\ldots a_i} 
\right)
\delta \phi^{A,m} d^{n-1}v= 0 \ .
\label{VAR.4}
\ee
As the $\delta \phi^{A,m}$'s are arbitrary 
we conclude that
\be
\sum_{i=0}^{k-m-1}(-1)^i\partial_{a_1}\ldots \partial_{a_i}\left(\frac{\partial Y^1}{\partial
\phi^{A,m}_{a_1\ldots a_i}} -R_{A,m+1}^{a_1\ldots a_i} 
\right)= 0 \ .
\label{VAR.5}
\ee
Let $\hat E^\lambda$ be the Noether current \eq{E.1} corresponding to
the Lagrange function $\hat {\cal L}$, as in \eq{VAR.0}. For our
purposes it is sufficient to consider vector fields $X^\lambda$ which
are transverse to $\Sigma$. We can choose a coordinate system in a
neighbourhood of $\Sigma$ so that $\Sigma$ is given by the equation
$x^1=0$, and moreover  $X^\lambda \partial_\lambda =
\partial_1$. From the definition of $\hat E^\lambda$ and $E^\lambda$ we
obtain
\begin{eqnarray*}
\hat E^1 & = & E^1 + \sum_{m=0}^{k-1}\sum_{i=0}^{k-m-1}
R_{A,m+1}^{a_1\ldots a_i}\phi^{A,m+1}_{a_1\ldots a_i} 
- \partial_\mu Y^\mu \\
& = &   E^1 + \sum_{m=0}^{k-1}\sum_{i=0}^{k-m-1}
\left(R_{A,m+1}^{a_1\ldots a_i}-\frac{\partial
Y^1}{\partial\phi^{A,m}_{a_1\ldots a_i}}\right) \phi^{A,m+1}_{a_1\ldots a_i}
- \partial_a Y^a\ .
\end{eqnarray*}
It follows that
\begin{eqnarray}
\int_\Sigma \hat E^\lambda dS_\lambda &-& 
\int_\Sigma  E^\lambda dS_\lambda \nonumber \\
&=&  \int_\Sigma
\sum_{m=0}^{k-1}\sum_{i=0}^{k-m-1}(-1)^i\partial_{a_1}\ldots
\partial_{a_i}\left(R_{A,m+1}^{a_1\ldots a_i}-\frac{\partial 
Y^1}{\partial\phi^{A,m}_{a_1\ldots a_i}}\right) \phi^{A,m+1}d^{n-1}v
\nonumber \\
&&
\label{VAR.6}  + \int_{\partial\Sigma} \left(Y^a -
\sum_{m=0}^{k-1} \sum_{i=0}^{k-m-2}  \phi^{A,m+1}_{a_1\ldots a_i} 
 \times \right. \\
&&\left. \sum_{\ell=0}^{k-m-i-2}(-1)^\ell
\partial_{b_1}\ldots\partial_{b_\ell} 
\left(R_{A,m+1}^{a_1\ldots a_ib_1\ldots b_\ell a}-\frac{\partial
Y^1}{\partial\phi^{A,m}_{a_1\ldots a_ib_1\ldots b_\ell
a}}\right)\right)
X^\lambda dS_{\lambda a} \ . \nonumber
\end{eqnarray}
The integral over $\Sigma$ in the right hand side of this last
equation vanishes by \eq{VAR.5}, which establishes our claim that the
Noether charge of $\Sigma$, defined as $\int_\Sigma  E^\lambda
dS_\lambda$, changes by a boundary integral under the change \eq{VAR.0}
of the Lagrange function.


\providecommand{\bysame}{\leavevmode\hbox to3em{\hrulefill}\thinspace}

\end{document}